\documentclass[twocolumn,superscriptaddress,prb,amsmath,amssymb]{revtex4}

\usepackage{graphics}
\usepackage{graphicx}
\usepackage{dcolumn}
\usepackage{bm}

\def\vp{\varphi}

\def\al{\alpha}

\def\eps{\epsilon}

\def\om{\omega}

\def\be{\begin{equation}}

\def\ee{\end{equation}}

\def\bea{\begin{eqnarray}}

\def\eea{\end{eqnarray}}

\def\bc{\begin{center}}

\def\ec{\end{center}}

\def\ra{\rightarrow}

\def\ov{\over}

\def\nonum{\nonumber}

\def\la{\langle}

\def\ra{\rangle}

\def\up{\uparrow}

\def\dn{\downarrow}

\begin{document}

\title{Bosonization approach to charge and spin dynamics of 1D fermions  \\ with band-curvature }

\author{Sofian Teber}

\email{sofian.teber@grenoble.cnrs.fr}

\affiliation{The Abdus Salam ICTP, Strada Costiera 11, 34014, Trieste, Italy}

\affiliation{Institut NEEL, CNRS and Universit\'e Joseph Fourier, BP 166, 38042, ÊGrenoble, France}

\date{\today}

\begin{abstract}
We consider one-dimensional (1D) spin-$1/2$ fermions in a clean quantum wire, with forward scattering interactions and a non-linear single-particle spectrum, $\xi_k = v|k| + k^2/2m$ where $v$ is the Fermi velocity and $1/m$ is the band-curvature. We calculate the dynamical structure factor (DSF) of the model at small wave-vector $q$ with the help of the bosonization technique. For spinless fermions, we show that, starting from the single-parametric spectrum: $\om = u |q|$, bosonization emulates the 2-parametric excitation spectrum: $\om = u |q| \pm q^2/2m^*$, where $m^*$ decreases with increasing repulsive interactions. Moreover, away from the excitation-cone, {\it i.e.} $\om \gg u |q|$, bosonization yields the 2-pair excitation continuum of the DSF. For spinful fermions,  we show that the spin-charge coupling (SCC) due to band-curvature affects charge and spin DSF in an asymmetric way. For the charge DSF, SCC manifests as a two-peak structure: a charge peak at $\om = u_\rho |q|$ but also a spin peak at $\om = u_\sigma |q|$, as charge fluctuations may decay via chargeless spin-singlet excitations. For the magnetic DSF, SCC manifests as a continuous transfer of magnetic spectral weight to frequencies $\om > u_\sigma |q|$, as spin fluctuations decay via pairs of chargeless spin and spinless charge-neutral excitations.
\end{abstract}

\maketitle


\section{Introduction}
\label{Introduction}

Correlation functions of one-dimensional (1D) fermions in quantum wires are conveniently calculated within the {\it Tomonaga-Luttinger} (TL) {\it model} and with the help of the bosonization technique. The TL model assumes the linearity of the single-particle spectrum with respect to momentum:
\be
\xi_k = v(|k| - k_F)~\quad~{\mathrm{(Tomonaga-Luttinger)}},
\label{spectrum_Luttinger}
\ee
where $v$ is the Fermi velocity and $k_F$ the Fermi momentum, and interactions of the forward scattering type, so-called $g_2$ and $g_4$ processes in $g$-ology. The forward scattering nature of the interactions implies that the system has no spectral gap whereas the linearity of the spectrum yields Lorentz invariance. These assumptions make the model exactly soluble by mapping the interacting fermions onto free bosonic excitations or plasmons [\onlinecite{Tomonaga,Luttinger,Schick,Bosonization,Haldane}]. As a consequence, {\it e.g.} transport, properties of TL liquids are easily accessed with the help of bosonization, see the recent monographs~[\onlinecite{Book:Giamarchi,Book:GNT}].

Even though the TL model allows a standard description of fermions in quantum wires, as Landau theory of Fermi liquids allows a standard description of 3D systems, {\it cf.} the monograph [\onlinecite{Book:Nozieres}], both assumptions of relativistic single-particle spectrum and forward scattering interactions often constitute an over simplified description of 1D fermionic systems. As a matter of fact, fermions hopping with an amplitude $t$ on a 1D lattice of parameter $a$  have a single-particle spectrum given by:
\bea
\xi_k = -2 t \cos(ka)~\quad~{\mathrm{(Hubbard)}}.
\label{spectrum_Hubbard}
\eea
Adding interactions between such fermions leads to the {\it 1D Hubbard model}, see the monograph~[\onlinecite{Book:Essler}], one of the most fundamental model of solid-state physics. In the absence of disorder and away from $1/2-$filling we may assume that the system has no spectral gap, {\it e.g.} is not in a Mott insulating phase. One may then focus on the low-energy properties of this model by expanding Eq.~(\ref{spectrum_Hubbard}) around the Fermi points $\pm k_F$ where the $+$ sign refers to right-movers and the $-$ sign to left-movers. In the lowest order, this yields the TL model, Eq.~(\ref{spectrum_Luttinger}), with $v = 2 t a \sin(k_Fa)$. Because of the exact solubility of the TL model the low-energy properties of the Hubbard model are thus known exactly. On the other hand, except for the case of free fermions which is again exactly soluble, less is known about the high-energy properties of the Hubbard model. Considerable work has been devoted to exact studies related to the nature of its ground-state and its low-energy excitations. This has been done with the help of the Bethe {\it Ansatz}, {\it cf.} [\onlinecite{Book:Essler,Lectures:Andrei}] for reviews, which allows a fully non-perturbative handling of the non-linear spectrum, Eq.~(\ref{spectrum_Hubbard}), and interaction effects. This technique is however extremely difficult to generalize to the calculation of correlation functions. Formidable efforts are made to achieve this task but they are limited at present to the $1/2-$filled case [\onlinecite{Essler_FF}] and to the XXZ Heisenberg spin-chain, see [\onlinecite{Kitanine}] for a review and Ref. [\onlinecite{Affleck}] for an application to the present problem (the XXZ spin-chain is equivalent to spinless fermions, {\it cf.} Ref.~[\onlinecite{Lectures:Affleck}] for a review on spin-chains and the Hubbard model).

In this contribution we focus on a different approach to the high-energy properties of strongly correlated 1D fermions. This amounts to start from the TL model, assume forward scattering interactions and introduce non-linear corrections to the linear single-particle spectrum. Restricting ourselves to the lowest order correction yields:
\be
\xi_k = v(|k|-k_F) + (|k| - k_F)^2/2m + \mathcal{O}((|k| - k_F)^3),
\label{non_linear_spectrum}
\ee
where $1/m = 2 t a^2 \cos(k_F a)$ is the band-curvature or, equivalently, $m$ is the band-mass. For Galilean invariant systems, there are no higher order corrections than the band-curvature. On the other hand, for lattice fermions, Eq.~(\ref{non_linear_spectrum}) neglects an infinite number of irrelevant (in the renormalization group sense) terms.  This means that, with respect to the Hubbard model Eq.~(\ref{spectrum_Hubbard}), the model defined by Eq.~(\ref{non_linear_spectrum}) only takes into account the most relevant of these terms which, below half-filling, corresponds to  the curvature of the band-spectrum. Such a simplified model then allows a non-perturbative treatment of band-curvature and to extract information about dynamical correlation functions of interacting 1D fermions {\it beyond} the TL model and {\it towards} the Hubbard model. 

Historically, such a generalization of the Tomonaga model has first been considered by Michael Schick~[\onlinecite{Schick}], see also~[\onlinecite{Haldane,BMN}]. More recently, this model has been considered in relation with the damping of 1D plasmons, Refs.~[\onlinecite{Samokhin,PK,Glazman:DSF_2006,Affleck}], and the drag resistivity due to forward scattering between quantum wires, Ref.~[\onlinecite{Glazman:DSF_2003}].  Most of these studies focus on the density-density correlation function in the case of spinless fermions, see Refs.~[\onlinecite{Glazman:DSF_2003,Rozhkov,PK,Glazman:DSF_2006,Affleck,ST}] or the equivalent XXZ Heisenberg spin-chain, see Refs.~[\onlinecite{Affleck,Glazman:DSF_2006}]. Various techniques have been considered to deal with this problem; mainly: fermionic [\onlinecite{Glazman:DSF_2003,Rozhkov,PK,Glazman:DSF_2006,ST}], field theory [\onlinecite{Affleck}], Bethe Ansatz  [\onlinecite{Affleck}] and numerical (Density-Matrix Renormalization Group) [\onlinecite{Affleck}]. Non-perturbative approaches to band-curvature may be found in [\onlinecite{Glazman:DSF_2003,Glazman:DSF_2006,Affleck}]. In this paper we will focus on a field theory (bosonization) approach to the computation of density-density 
correlation functions and consider the case of fermions with spin.

The paper is structured as follows. 
In Sec.~\ref{Motivations} we will review some known results (for free as well as interacting 
[\onlinecite{Glazman:DSF_2006,Affleck}]) fermions and motivate the need for a bosonization approach. 
In Sec.~\ref{SL_section} the case of spinless fermions will be considered and the results compared with the ones found in the literature, 
cf. [\onlinecite{Glazman:DSF_2006,Affleck}].  
In Sec.~\ref{SF_section} the case of spin$-1/2$ fermions will be considered following Ref.~[\onlinecite{BMN}].  
In Sec.~\ref{Conclusion} we will summarize our results and conclude. 


\section{Motivations for a bosonization approach}
\label{Motivations}

The main object we will consider in the following is the density-density correlation function or polarization operator:
\be
\Pi(x,\tau) = \la \rho(x,\tau)\rho(0,0) \ra,
\nonum
\ee
where $\rho$ is the fermion density measured relative to a neutralizing jellium. 

\begin{figure}
\includegraphics[width=4.5cm,height=4cm]{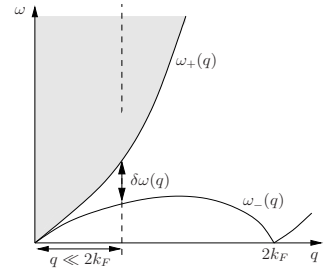}
\caption{ \label{Spectrum} Schematic view on the spectrum of excitations of 1D spinless fermions. In the case of free fermions, a single-pair excitation continuum, $S^0$, lies between the $\om_-$ and $\om_+$ branches. In the presence of forward scattering interactions a multi-pair excitation continuum ($S^{(2)}$ at the $2-$pair level) spreads above the $\om_+$ branch and a divergent 
spectral weight appears along the $\om_-$branch, {\it cf.} Fig.\ref{DSF}.}
\end{figure}

For {\it free-fermions} this correlation function is known exactly with band-curvature. It is simply the polarization bubble, i.e. the convolution of two free fermionic Green's functions. In Fourier space it reads:
\be
\Pi^0(i\om,q)= { m \ov \pi q}
\ln{ \left[{(i \om)^2 - (\om_-^0)^2 \ov (i\om)^2 - (\om_+^0)^2} \right]},
\label{PolarNL}
\ee
where the upper index refers to free-fermions and we have taken into account the spin degeneracy. In Eq.~(\ref{PolarNL}), the $2-$parametric family appears:
\be
\om_\pm^0 (q) = vq \pm {q^2 \ov 2m}.
\label{2parametric}
\ee
In the (particle-hole) excitation spectrum of the system, this family bounds the region of decay of the coherent bosonic excitation, the plasmon which corresponds to a coherent particle-hole pair, into the incoherent continuum of single particle-hole pairs, see Fig.~\ref{Spectrum}. This is best revealed by looking at the dynamic structure factor (DSF) of the fermions which corresponds to the dissipative part of the retarded polarization operator:
\be
S(\om, q) = - \Im \Pi^R (\om,q).
\label{DSF_definition}
\ee
For free fermions, Eqs.~(\ref{PolarNL}) and (\ref{DSF_definition}) yield (at $T=0$):
\bea
&&S^0(\om,q) = { m \ov |q|} \left[ \theta [{mv \ov q}(\om - \om_-^0)] - \theta [{mv \ov q}(\om - \om_+^0)] \right]
\nonum \\
&&- \{ \om \rightarrow -\om\},
\label{DSFNL}
\eea
which, for a fixed momentum $q$, is non-zero and independent on frequency in the range: $\om_-^0 < \om < \om_+^0$. In particular, at the plasmon frequency, Eq.~(\ref{DSFNL}) yields a finite result:
\be
S^0(\om = vq, q) = {m \ov |q|}.
\label{plasmon_LT_BC}
\ee
Moreover, the spectral width of the DSF yields the inverse life-time of the bosonic excitations:
\be
\delta \om^0 = | \om_+^0 - \om_-^0| = {q^2 \ov m}.
\label{plasmon_LT2_BC}
\ee
\begin{figure}
\includegraphics[width=6.5cm,height=3.5cm]{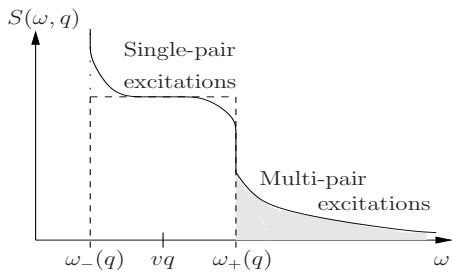}
\caption{ \label{DSF} Schematic view on the dynamical structure factor of interacting spinless fermions as a function of frequency $\om$ for a given momentum $q$, {\it {\it e.g.}} the cut at $q \ll 2k_F$ of Fig.~\ref{Spectrum} (at $T=0$). The single-pair excitation continuum lies between the $\om_-$ and $\om_+$ branches. As known from Refs. [\onlinecite{Glazman:DSF_2006,Affleck}], the spectral weight diverges at the lower-edge $\omega_-$ and crosses over smoothly to a multi-pair continuum at the upper edge $\omega_+$. }
\end{figure}

On the other hand, expanding Eq.~(\ref{PolarNL}) in $1/m$ yields:
\bea
&&\Pi^0 (i\om,q)
\label{FF_PO} \\
&&= {2 \ov \pi v}~\sum_{n = 0}^{\infty}~{1 \ov 2n+1}~\left( {q^2 \ov m} \right)^{2n} {(vq)^{2n+2} \ov [(i\om)^2 - (vq)^2]^{2n+1}},
\nonum
\eea
which reveals a basic difference with respect to the exact result of Eq.~(\ref{PolarNL}): order by order the dissipative part is singular at the plasmon frequency: $\om = v|q|$, with $\delta-$function singularities. In particular, within the TL approximation ($m \rightarrow \infty$):
\be
S^{0}_{TL}(\om,q) = {|q|}~\delta[\om-v|q|],
\label{plasmon_LT_TL}
\ee
which implies that plasmons have an infinite life-time, {\it i.e.} are free. The divergency of the DSF, in the limit $m \rightarrow \infty$, agrees with Eq.~(\ref{plasmon_LT_BC}) but the nature of the divergency is different according to the order with which the limits $m \rightarrow \infty$ and $\om \rightarrow v q$ are taken. The fact that these limits do not commute witness the non-trivial role of the irrelevant band-curvature correction. Indeed, the latter cures the TL $\delta-$function singularities, {\it cf.} Eq.~(\ref{plasmon_LT_BC}), providing the exact line-shape of the free-fermion DSF, {\it cf.} Eq.~(\ref{DSFNL}), and the finite life-time of the elementary excitations, {\it cf.} Eq.~(\ref{plasmon_LT2_BC}). Moreover, all these results are non-perturbative in band-curvature as they require summing over an infinite number of singular terms.

The case of {\it interacting fermions} has been the subject of the recent literature. The results are summarized schematically on Fig.~\ref{DSF}. The latter shows that the interplay between band-curvature and interactions for the DSF is two-fold. On the one hand, at high frequencies ($\om \gg vq$), this interplay yields a multi-pair continuum to the DSF. The $2-$pair continuum was first derived in Ref.  [\onlinecite{Glazman:DSF_2003}] and subsequently in Refs. [\onlinecite{Affleck,ST}]. In Refs. [\onlinecite{Glazman:DSF_2003,ST}] fermionic techniques were used to derive the $2-$pair continuum whereas 
Ref.  [\onlinecite{Affleck}] derived it with the help of the bosonization technique. This high frequency continuum may be accessed by perturbation theory in curvature and interactions. 
On the other hand, at low frequencies ($\om \approx vq$), a (power-law) divergency of the spectral weight appears at the lower-edge $\omega_-$ and a smooth crossover to the multi-pair continuum takes place at the upper edge $\omega_+$, [\onlinecite{Glazman:DSF_2006,Affleck}] . Moreover, the width of the single-pair continuum was shown to be affected by interactions [\onlinecite{Glazman:DSF_2003,Affleck}]:
\be
\delta \om \propto {q^2 \ov m^*},
\label{plasmon_width}
\ee
where $m^*$ gets smaller with increasing repulsive interactions, {\it i.e.} the plasmon peak becomes broader. Contrary to the high-frequency part the low-frequency regime requires a non-perturbative treatment of band-curvature. This was already the case for free fermions. Adding interactions of course considerably complicates the problem. This was achieved in Ref.~[\onlinecite{Glazman:DSF_2003,Glazman:DSF_2006}] by fermionic techniques and in  Ref.~[\onlinecite{Affleck}] by Bethe Ansatz and numerical simulation. As can be seen from Fig.~\ref{DSF}  the free-fermion box-like line-shape is therefore significantly modified by interactions, both at low and high frequencies.

In this frame, our goal will be to calculate the DSF (at small $q$, {\it i.e.} the $2k_F-$part is not considered) with the help of the bosonization technique. We will start from a Hubbard chain below $1/2-$filling and show in detail how the bosonization can be implemented when taking into account band-curvature corrections to the single-particle spectrum.  Our first motivation for using bosonization comes from the fact that it allows to take forward scattering interactions among the fermions to all orders even though it is essentially perturbative in band-curvature.  Following the recent Ref.  [\onlinecite{Affleck}] we will reconsider the case of spinless fermions. The exact results available in this case will allow us to discuss the efficiency and limitation of this technique towards a possible re-summation of band-curvature corrections. Our second motivation comes from the fact that bosonization allows to take quite straightforwardly into account the spin-$1/2$ of the fermions. We will focus on the highly non-trivial coupling between spin and charge degrees of freedom due to band-curvature. Within bosonization, this was first noticed in Ref.~[\onlinecite{BMN}] that we shall follow (correlation functions were not the main issue of Ref.~[\onlinecite{BMN}]). A spin-charge coupling due to band-spectrum non-linearities was also reported in Ref.~[\onlinecite{Lectures:Andrei}] within a Bethe Ansatz approach to the 1D Hubbard model (correlation functions are not yet computable with this technique). As we will detail below, field and lattice theories agree qualitatively. On this basis, we will explore, with the help of the bosonization technique, the effect of this coupling on the charge and spin density correlation functions of the model. At this point, the reader which is interested more in our results than the derivations may refer directly to Sec.~\ref{Conclusion}.


\section{Spinless fermions}
\label{SL_section}

\subsection{Model}
\label{SL_model}

In this Section, we consider spinless fermions and give some details of the standard bosonization procedure. Starting with free fermions their Hamiltonian reads:
\be
H = \sum_k~\xi_k~:c^\dagger_k c_k:,
\nonum
\ee
where $c$ and $c^\dagger$ are the annihilation and creation operators for the fermions, $\xi_k$ is given by Eq.~(\ref{non_linear_spectrum})  and $::$ refers to normal ordering. The bosonization procedure is implemented by first focusing on the low-energy sector near the Fermi points $\pm k_F$:
\be
c(x) \sim \psi_+(x) e^{-ik_Fx} + \psi_-(x) e^{+ik_Fx},
\nonum
\ee
where $\psi_\pm$ are slow chiral fields. The free fermion Hamiltonian becomes:
\bea
H &&=i v~\int {{dx}}~\left[ :\psi_+^\dagger  \partial_x \psi_+: - :\psi_-^\dagger \partial_x \psi_-: \right]
\nonum \\
&&- {1 \ov 2m}~\int {{dx}}~\left[  :\psi_+^\dagger \partial_{xx} \psi_+: + :\psi_-^\dagger \partial_{xx} \psi_-: \right].
\label{free_fermion_H}
\eea
The first term corresponds to the Dirac part related to the linear spectrum. The second term corresponds to the curvature correction. One may then go to the bosonic representation with the help of the following identity:
\be
\psi_\pm(x) = {1 \ov \sqrt{2 \pi}}~:\exp \left[ \pm i \sqrt{4 \pi}~\vp_\pm(x) \right]:,
\label{vertex_operator}
\ee
relating the chiral fermionic fields, $\psi_\pm$, to the chiral bosonic fields, $\vp_\pm$. The latter may be expressed as a function of the total phase field $\vp$ and the momentum $\Pi_\vp$ as follows:
\be
\vp_\pm = {1 \ov 2}~\left(\vp \mp \int_{-\infty}^{x} {{dx'}}~\Pi_\vp(x') \right).
\label{phase}
\ee
The normal-ordering of vertex operators is conveniently taken into account with the help of a point-splitting technique. This yields the following operator-product expansions ($\eta = \pm$):
\bea
&&\psi_\eta^\dagger(x) \psi_\eta(x) = - {1 \ov \sqrt{\pi}}~\partial_x \vp_\eta(x),
\nonum \\
&&\psi_\eta^\dagger(x) \partial_x \psi_\eta(x) = - i~{\mathrm{sgn}}(\eta)~:(\partial_x \vp_\eta(x))^2:,
\nonum \\
&&\psi_\eta^\dagger(x) \partial_{xx} \psi_\eta(x) = {4 \sqrt{\pi} \ov 3}~:(\partial_x \vp_\eta(x))^3:,
\label{ope}
\eea
where the terms which vanish upon integration over space or by symmetry considerations have been neglected. The last line in Eq.~(\ref{ope}) displays the non-linearity of the bosonization arising from band-curvature. The fermionic Hamiltonian then takes over the following form:
\bea
H &&= v~\int {{dx}} \left[ (\partial_x \vp_+)^2 + (\partial_x \vp_-)^2 \right]
\nonum \\
&&- {2 \sqrt{\pi} \ov 3 m}~\int {{dx}}~\left[ (\partial_x \vp_+)^3 + (\partial_x \vp_-)^3 \right],
\label{free_fermion_H_chiral}
\eea
with the densities:
\be
\rho_+(x) = - {1 \ov \sqrt{\pi}}~\partial_x \vp_+(x),~~\rho_-(x) = - {1 \ov \sqrt{\pi}}~\partial_x \vp_-(x).
\nonum
\ee
The fact that the curvature part is cubic in densities is related to the energy of the Fermi sea for a quadratic spectrum in 1D. Moreover, such cubic terms lead to a violation of the particle-hole symmetry, $\vp_\pm \rightarrow - \vp_\pm$, again a consequence of the non-linearity of the spectrum away from half-filling. Finally, notice that the curvature term in Eq.~(\ref{free_fermion_H_chiral}) agrees with the one found in the literature [\onlinecite{Haldane,Samokhin,Affleck}], see Appendix~\ref{A:Notations} for conventions on notations.

In terms of the total phase and momentum this yields:
\bea
H &&= {v \ov 2}~\int {{dx}} \left[ \Pi_\vp^2 + (\partial_x \vp)^2 \right]
\nonum \\
&&-{\sqrt{\pi} \ov 6 m}~\int {{dx}}~\partial_x \vp~\left[ 3 \Pi_\vp^2 + (\partial_x \vp)^2 \right].
\label{free_fermion_H_mixed}
\eea
According to phenomenological bosonization the added terms, $(\partial_x \vp)^3$ and $\partial_x \vp ~\Pi_\vp^2$, are the lowest-order particle-hole symmetry violating terms compatible with both the $U(1)$ invariance of the Hamiltonian, {\it i.e.} $\vp \rightarrow \vp + \al$ where $\al$ generates the translations, and the space-time symmetries of the problem: $\vp$ is an odd function of $x$ and an even function of $t$:
\be
\rho = -{1 \ov \sqrt{\pi}}~\partial_x \vp, ~~~ j = {1 \ov \sqrt{\pi}}~\partial_t \vp,
\label{density_spinless}
\ee
where $\rho$ is the total density (the smooth part of the density close to $q=0$; the $2k_F$ part has been neglected), $j$ the total current of the fermions and $\Pi_\vp \sim \partial_t \vp$, see Eq.~(\ref{CM}) below. Notice also, from Eq.~(\ref{density_spinless}), that the original (spinless) fermions correspond to kinks in $\vp$.

To include forward scattering interactions we follow the usual prescription of re-scaling the fields:
\be
\Pi_\vp \rightarrow \sqrt{\gamma_\rho}~\Pi_\vp,~~\vp \rightarrow {1 \ov \sqrt{\gamma_\rho}}~\vp,~~v \rightarrow u_\rho,
\nonum
\ee
in the Tomonaga-Luttinger part of the Hamiltonian, where:
\bea
\gamma_\rho &&= \sqrt{{1 + y_{4,c}/ 2 - y_{2,c} / 2 \ov 1 + y_{4,c} / 2 + y_{2,c} / 2}},
\nonum \\
u_\rho &&= v \sqrt{(1 + y_{4,c}/2)^2 - (y_{2,c}/2)^2},
\nonum
\eea
and $y_{i,c} \equiv g_{i,c} / \pi v$ is the dimensionless coupling constant in the charge sector. In the following we take: $y_{2,c} = y_{4,c} = y_c$ which yields:
\be
\gamma_\rho = {1 \ov \sqrt{1 + y_c}},~~u_\rho = v~\sqrt{1 + y_c}.
\label{parameters_charge}
\ee
The interactions affect the free fermion Hamiltonian density as:
\be
\mathcal{H} = {u_\rho \ov 2}~\left[ \gamma_\rho~\Pi_\vp^2 + {1 \ov \gamma_\rho}~(\partial_x \vp)^2 \right] - {\sqrt{\pi} \ov 6 m}~\partial_x \vp~\left[ 3 \Pi_\vp^2 + (\partial_x \vp)^2 \right].
\label{fermion_H_mixed}
\ee
To proceed further, we find it more convenient to go to the Lagrangian representation, $\Pi_\vp \rightarrow \partial_x \theta$, which yields the following Euclidean action:
\bea
S_{E} = \int {{d\tau dx}}~\left[ {u_\rho \gamma_\rho \ov 2}~[1 - {6 \ov m' u_\rho \gamma_\rho}~\partial_x \vp]~(\partial_x \theta)^2 \right.
\nonum \\
\left. -i \partial_x \theta \partial_\tau \vp + {u_\rho \ov 2 \gamma_\rho}~(\partial_x \vp)^2 - {1 \ov m'}~(\partial_x \vp)^3 \right],
\nonum
\eea
where $m'=6m/\sqrt{\pi}$ and $\tau=it$ is the imaginary time. Integrating over the dual phase field $\theta$ yields:
\bea
\mathcal{L}_E [\vp] &&= {1 \ov 2 \gamma_\rho u_\rho}~{1 \ov {1 - {6 \ov m' u_\rho \gamma_\rho} \partial_x \vp}}~(\partial_\tau \vp)^2 + {u_\rho \ov 2 \gamma_\rho}~(\partial_x \vp)^2
\nonum \\
&&- {1 \ov m'}~(\partial_x \vp)^3.
\nonum
\eea
The non-trivial denominator comes from the conjugated momentum which reads:
\be
\Pi_\vp = { \partial_t \vp \ov u_\rho \gamma_\rho~(1 - {6 \ov m' u_\rho \gamma_\rho} \partial_x \vp)}.
\label{CM}
\ee
Expanding the previous expression in the lowest meaningful order in $1/m$ yields:
\bea
\mathcal{L}_E [\vp] &&= {1 \ov 2 v} \left[ (\partial_\tau \vp)^2 + u_\rho^2~(\partial_x \vp)^2 \right]
\nonum \\
&&+ {1  \ov m' v^2}~\partial_x \vp \left[ 3~(\partial_\tau \vp)^2 - v^2~(\partial_x \vp)^2 \right],
\label{lagrangian}
\eea
where we have used the fact that: $\gamma_\rho u_\rho = v$, independent of interactions~\cite{Note:parameters}.

As an alternative to Eq.~(\ref{lagrangian}), we may work with the complete action:
\bea
\mathcal{L}_{E}[\vp,\theta] &&= {u_\rho \gamma_\rho \ov 2}~(\partial_x \theta)^2 + {u_\rho \ov 2 \gamma_\rho}~(\partial_x \vp)^2 -i \partial_x \theta \partial_\tau \vp
\nonum \\
&&- {3 \ov m'}~\partial_x \vp~(\partial_x \theta)^2  - {1 \ov m'}~(\partial_x \vp)^3,
\label{lagrangian2}
\eea
which does not require any expansion in $1/m$.

With the help of Eq.~(\ref{lagrangian}) or Eq.~(\ref{lagrangian2}) we wish to determine the effect of the irrelevant terms appearing due to curvature on the correlation functions of the model. In particular the long-distance part (close to $q=0$; the $2k_F-$part is neglected) of the polarization operator reads:
\be
\Pi(x,\tau) =  {2 \ov \pi}~\la~\partial_x \vp(x,\tau)~\partial_x \vp(0,0)~\ra,
\nonum
\ee
where we have taken into account the spin degeneracy.
From its expression in Fourier space all we need to compute is the phase correlator $\mathcal{D}$:
\be
\Pi(i\om,q) = - {2 q^2 \ov \pi}~\mathcal{D}(i\om,q), ~~~~~\mathcal{D}(i\om,q) = \la |\vp(i\om,q)|^2 \ra.
\label{pi_definition}
\ee
Curvature terms make the actions non-Gaussian a signature of the fact that they induce interactions between the bosonic excitations. We will treat them in perturbation theory in $1/m$. Notice that for Eq.~(\ref{lagrangian}) curvature generates an infinite number of irrelevant terms and we have taken into account only the lowest order terms. This means that our treatment of curvature is essentially perturbative (even if we manage to re-sum all terms generated by Eq.~(\ref{lagrangian})). At second-order, both models of Eq.~(\ref{lagrangian}) and Eq.~(\ref{lagrangian2}) are equivalent as will be shown in more details below.


\subsection{Dynamical structure factor}
\label{SL_CF}

We now compute $\mathcal{D}(i\om,q)$ via bosonization, {\it i.e.} in perturbation theory in $1/m$. From Eq.~(\ref{lagrangian}) the zeroth order reads:
\be
\mathcal{D}^{(0)}(i\om,q) = - {u_\rho \gamma_\rho \ov (i\om)^2 - (u_\rho q)^2},
\label{zero_green}
\ee
from which:
\be
\Pi^{(0)}(i\om,q) = {2 \gamma_\rho \ov \pi u_\rho}~{(u_\rho q)^2 \ov (i\om)^2 - (u_\rho q)^2},
\label{Pi_TL}
\ee
which agrees with the zero order term of Eq.~(\ref{FF_PO}) in the free-fermion limit: $\gamma_\rho = 1$ and $u_\rho = v$.

\begin{figure}
\includegraphics[width=2cm,height=1cm]{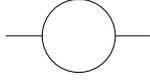}
\caption{ \label{Diagrams_vp_2} Second order diagram of the bosonic field theory for the $\vp-$Lagrangian of Eq.~(\ref{lagrangian}) for spinless fermions.}
\end{figure}

In the Euclidean action the cubic perturbation reads:
\bea
&& S_E = {i \ov m' v^2} {1 \ov \beta^3} \sum_{\om_1,\om_2,\om_3} \int {{dq_1 dq_2 dq_3 \ov (2 \pi)^3}} \delta[1+2+3]
\nonum \\
&&\left[3~q_1 \om_2 \om_3 - v^2~q_1 q_2 q_3 \right]  \vp_{\om_1}(q_1) \vp_{\om_2}(q_2) \vp_{\om_3}(q_3).
\nonum
\eea
As will be discussed in the following the knowledge of the self-energy associated with this perturbation (at least in the lowest order) does not imply any meaningful re-summation of band-curvature effects at the level of the Dyson equation. We will therefore simply calculate corrections to the bare Bose propagator order by order: 
\be
\mathcal{D}(i\om,q) = \mathcal{D}^{(0)}(i\om,q) + \mathcal{D}^{(2)}(i\om,q) +~..., 
\ee
and similarly for the polarization operator. The lowest order correction is of the second order in band-curvature. It reads:
\be
\mathcal{D}^{(2)}(i\om,q) = \mathcal{D}^{(0)}(i\om,q)~\Sigma(i\om,q)~\mathcal{D}^{(0)}(i\om,q),
\nonum
\ee
where $\Sigma(i\om,q)$ is the second-order self-energy part and is used without any re-summation implied.  From the expression of the free bosonic Green function, Eq.~(\ref{zero_green}), it will allow us to compute the second-order correction to the polarization operator, Eq.~(\ref{pi_definition}):
\be
\Pi^{(2)}(i\om,q) = - { 2 \gamma_\rho^2 \ov \pi}~{(u_\rho q)^2 \ov [(i\om)^2 -(u_\rho q)^2]^2}~\Sigma(i\om,q).
\label{pi_definition2}
\ee
For the $\vp-$Lagrangian the diagrammatic theory assigns a solid line for the bare propagator of Eq.~(\ref{zero_green}) and the second-order self-energy part is displayed on Fig.~\ref{Diagrams_vp_2}. To compute this self-energy part one has to be careful in Wick ordering the 8-point correlation functions generated by the cubic action at second order. This care is due to the fact that, in the $\vp-$representation, the vertices carry products of frequency and momenta in a non-symmetric way which complicates the calculation of the combinatorial factor. The latter is crucial as we will see in the following.

As an alternative to the $\vp-$representation we may work in the $\vp,\theta-$representation. As we have noticed in Sec.~\ref{SL_model} the $\vp-$Lagrangian of Eq.~(\ref{lagrangian}) is derived by neglecting an infinite number of irrelevant perturbations. On the other hand no such truncation appears in the $\vp,\theta-$Lagrangian of Eq.~(\ref{lagrangian2}). At the second order of the perturbation theory in curvature both approaches should yield the same results while at higher orders the $\vp,\theta-$Lagrangian has to be used. Moreover, in the $\vp,\theta-$representation the vertices carry products of momentum in a symmetric way which simplifies the Wick ordering at the expense of having more diagrams to evaluate. The latter can be seen from the Fourier transform of Eq.~(\ref{lagrangian2}) which yields the following perturbations:
\bea
&& S_E = -{i \ov m'}~{1 \ov \beta^3}~\sum_{\om_1,\om_2,\om_3}~\int {{dq_1 dq_2 dq_3 \ov (2 \pi)^3}}~\delta[1+2+3]
\nonum \\
&& q_1 q_2 q_3~\left[3 \vp_{\om_1}(q_1) \theta_{\om_2}(q_2) \theta_{\om_3}(q_3)+ \vp_{\om_1}(q_1) \vp_{\om_2}(q_2) \vp_{\om_3}(q_3) \right].
\nonum
\eea
The quadratic part of Eq.~(\ref{lagrangian2}) yields the zero-order correlators:
\bea
&&\mathcal{D}^{(0)}_{\vp \vp}(i\om,q) = \la |\vp(i\om,q)|^2 \ra = - { u_\rho \gamma_\rho \ov (i\om)^2 - (u_\rho q)^2},
\nonum \\
&&\mathcal{D}^{(0)}_{\theta \theta}(i\om,q) = \la |\theta(i\om,q)|^2 \ra = -{u_\rho \ov \gamma_\rho}~{1 \ov (i\om)^2 - (u_\rho q)^2},
\nonum \\
&&\mathcal{D}^{(0)}_{\vp \theta}(i\om,q) = \la \vp(i\om,q) \theta(-i\om,-q) \ra =  {- i\om \ov q[(i\om)^2 -(u_\rho q)^2]},
\nonum \\
&&\mathcal{D}^{(0)}_{\theta \vp}(i\om,q) = \mathcal{D}^{(0)}_{\vp \theta}(i\om,q).
\label{bare_GF_spinless}
\eea
Notice that while the charge phase $\vp$ is related to particle-hole pairings, $\mathcal{D}^{(0)}_{\vp \vp} \sim \Pi^{(0)}$, its dual, $\theta$, is a "superconducting" phase and $\mathcal{D}^{(0)}_{\theta \theta}$ is related to particle-particle pairings. The fact that $\vp$ and $\theta$ appear on equal footing in Eq.~(\ref{lagrangian}) is related to the absence of long-range order (either of the charge-density wave type or the superconducting one) in 1D.

\begin{figure}
\includegraphics[width=8cm,height=3.5cm]{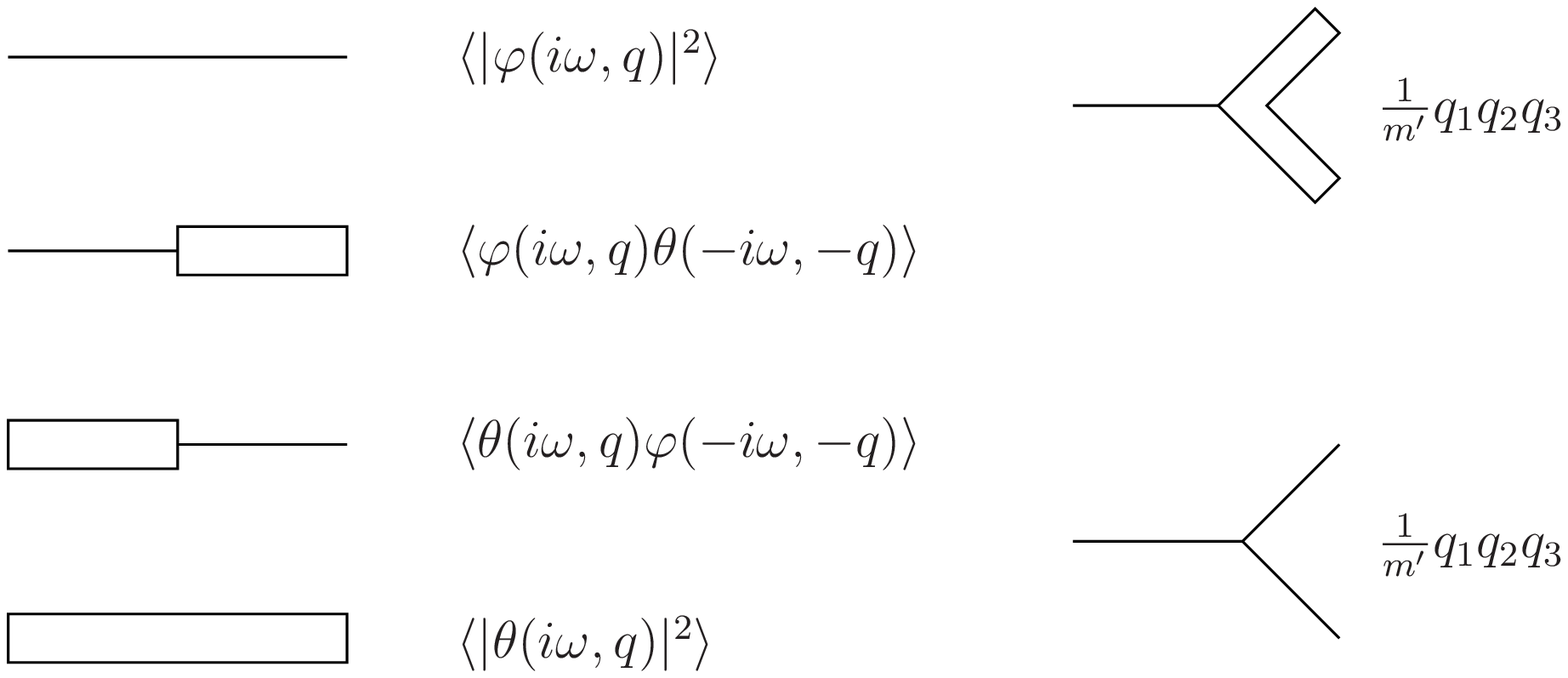}
\caption{ \label{Spinless_QFT} The quantum field theory associated with the $\vp,\theta-$Lagrangian of Eq.~(\ref{lagrangian2}) for spinless fermions.}
\end{figure}

The diagrammatic theory then assigns a solid line for the $\vp-$field and a double line for the $\theta-$field, see Fig.~\ref{Spinless_QFT}.
\begin{figure}
\includegraphics[width=8cm,height=7cm]{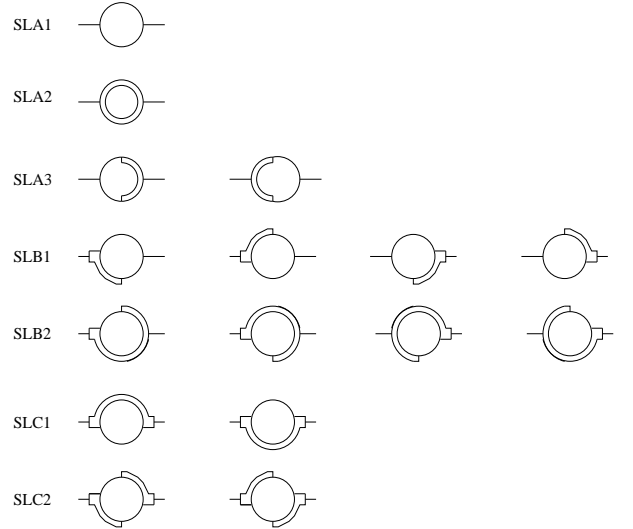}
\caption{ \label{Diagrams} Second order diagrams of the bosonic field theory for the $\vp,\theta-$Lagrangian of Eq.~(\ref{lagrangian2}) for spinless fermions. SL stands for: Spin Less.}
\end{figure}
Expanding the perturbation up to second order and Wick ordering the $8-$point correlation functions yields seven different types of diagrams represented on Fig.~\ref{Diagrams}. They are classified according to the number of $\theta-$wings attached to the self-energy parts. These diagrams, Fig.~\ref{Diagrams}, yield exactly the same self-energy contribution as the one of the $\vp-$representation, Fig.~\ref{Diagrams_vp_2}.

The final result for the small $q$ self-energy reads (at $T=0$ for simplicity):
\bea
&&\Im \Sigma_\rho^R(\om,q) =
\label{SE_SF} \\
&&~{\gamma_\rho^2 \pi \ov 96 u_\rho}~\left [ 1 + {3 \ov \gamma_\rho^2} \right]^2~\left( q^2 \ov m \right)^2~u_\rho q~\left[ \delta[\om - u_\rho q] - \delta[\om + u_\rho q] \right]
\nonum \\
&&+ {\gamma_\rho^2 \pi \ov 64 u_\rho}~\left[ 1 - {1 \ov \gamma_\rho^2} \right]^2~\left( q \ov m u_\rho \right)^2~\left[ \om^2 - (u_\rho q)^2 \right]~\mathcal{F}_\rho(\om,q),
\nonum
\eea
where we have restored $m' = 6m / \sqrt{\pi}$ and defined the kinematic factor:
\be
\mathcal{F}_\rho(\om,q) =  \theta[\om - u_\rho q] -  \theta[-\om - u_\rho q].
\label{F_SL}
\ee
%

\subsubsection{Multi-pair continuum}

Away from the plasmon-cone, $\om \gg u_\rho |q|$, Eq.~(\ref{SE_SF}) yields the small wave-vector, high-frequency part of the DSF:
\be
S_\rho^{(2)}(\om,q) = {( 1 -  \gamma_\rho^2)^2 \ov 32 u_\rho}~\left( q^2 \ov m \right)^2~{ 1 \ov \om^2 - (u_\rho q)^2 }~\mathcal{F}_\rho(\om,q),
\label{dsf_tail}
\ee
which is the leading contribution to the DSF in this frequency regime.
This tail is by now well known in the literature, cf. Refs.  [\onlinecite{Glazman:DSF_2003,Affleck,ST}], and describes the decay of a plasmon into two plasmon excitations as can be seen from the vertices of Fig.~\ref{Spinless_QFT}. In fermionic language the plasmon corresponds to a coherent particle-hole pair. As a consequence, the tail is termed the two-pair excitation continuum, see Figs.~\ref{Spectrum} and \ref{DSF}. Notice that the kinematic factor appearing in Eq.~(\ref{dsf_tail}), see also Eq.~(\ref{F_SL}), is due to the fact that, in 1D, low-energy particle-hole pairs cannot exist for arbitrarily small momentum.

\subsubsection{Single-pair continuum}

\begin{figure}
\includegraphics[width=6cm,height=1.5cm]{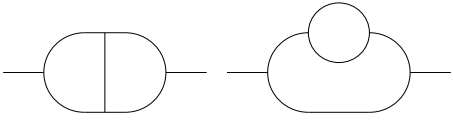}
\caption{ \label{Diagrams_vp_4} Fourth order diagrams of the bosonic field theory for the $\vp-$Lagrangian of spinless fermions.}
\end{figure}

Close to the plasmon-cone, $\om =  u_\rho |q|$,  the total polarization operator may be again conveniently re-casted in the Matsubara form: 
\be
\Pi(i\om,q) = \Pi^{(0)}(i\om,q) + \Pi^{(2)}(i\om,q) + ...,
\ee
where the dots refer to higher order band-curvature corrections, {\it cf.} the fourth order contributions of Fig.~\ref{Diagrams_vp_4}, and from Eqs.~(\ref{SE_SF}) and (\ref{pi_definition2}):
\be
\Pi^{(2)}(i\om,q) = {2 \ov 3 \pi u_\rho}~\left( {3+ \gamma_\rho^2 \ov 4} \right)^2~\left( q^2 \ov m \right)^2~{ (u_\rho q)^4 \ov [(i\om)^2 - (u_\rho q)^2]^3}.
\label{dsf_core}
\ee
We now perform a re-summation of band-curvature corrections from the second-order self-energy part. For $q \ll k_F$, this yields new poles for the dressed polarization operator:
\be
\om_\pm(q) = u_\rho |q| \pm \sqrt{2 \ov 3}~{3 +\gamma_\rho^2 \ov 4 \sqrt{\gamma_\rho}}~{q^2 \ov 2 m}.
\label{Poles}
\ee
We therefore recover the 2-parametric structure of the single-pair continuum. Moreover, up to a numerical prefactor of the order of unity, the width:
\be
\delta \om \propto {q^2 \ov m^*},~~~~m^* =  { 4 \sqrt{\gamma_\rho} \ov 3 + \gamma_\rho^2} m,
\label{effective_m}
\ee
increases with increasing repulsive interactions. This agrees qualitatively with the exact results of Refs.  [\onlinecite{Glazman:DSF_2006,Affleck}].

On the other hand, the precise line-shape of the DSF in the single-pair continuum (divergency on the lower edge and crossover to the multi-pair continuum at the upper edge, see Fig.~\ref{DSF}) cannot be accessed, even qualitatively, with the help of the second order self-energy. Indeed, 
such a partial re-summation does not cure the $\delta-$function singularities at the new frequencies of Eq.~(\ref{Poles}). Higher order corrections, some of which are represented on Fig.  \ref{Diagrams_vp_4}, have to be taken into account in order to develop a more efficient re-summation scheme. This important issue is however beyond the scope of the present paper.

\section{spinful fermions}
\label{SF_section}

\subsection{Model and spin-charge coupling}
\label{SF_Model}

Following the previous sections we now move to spinful fermions by introducing an additional internal, spin-$1/2$, degree of freedom to the fields: $\sigma = \up, \dn$, so that:
\bea
\psi_{\pm, \sigma}(x) &&= {1 \ov \sqrt{2 \pi}}~:\exp \left[ \pm i \sqrt{4 \pi}~\vp_{\pm,\sigma}(x) \right]:,
\nonum \\
\vp_{\pm,\sigma} &&= {1 \ov 2}~\left(\vp_{\sigma} \mp \int_{-\infty}^{x} {{dx'}}~\Pi_{\sigma}(x') \right).
\nonum
\eea
This yields:
\bea
H &&= v~\sum_\sigma~\int {{dx}} \left[ (\partial_x \vp_{+,\sigma})^2 + (\partial_x \vp_{-,\sigma})^2 \right]
\nonum \\
&&- {2 \sqrt{\pi} \ov 3 m}~\int {{dx}}~\left[ (\partial_x \vp_{+,\sigma})^3 + (\partial_x \vp_{-,\sigma})^3 \right],
\nonum
\eea
with the densities:
\be
\rho_{\eta,\sigma}(x) = - {1 \ov \sqrt{\pi}}~\partial_x \vp_{\eta,\sigma}(x).
\nonum
\ee
We then introduce the charge and spin fields together with their canonically conjugated fields:
\bea
&&~~\vp = {1 \ov \sqrt{2}}~[\vp_\up + \vp_\dn], \qquad \sigma = {1 \ov \sqrt{2}}~[\vp_\up - \vp_\dn],
\nonum \\
&&\Pi_\vp = {1 \ov \sqrt{2}}~[\Pi_\up + \Pi_\dn], \qquad \Pi_\sigma = {1 \ov \sqrt{2}}~[\Pi_\up - \Pi_\dn].
\nonum
\eea
Following Ref.~[\onlinecite{BMN}], the Hamiltonian density then takes the following form:
\begin{subequations}
\label{hamiltonian_spinful}
\bea
&&\mathcal{H} ~~~= ~\mathcal{H}_{TL} + \mathcal{H}_C,
\label{total_hamiltonian_spinful} \\
&&\mathcal{H}_{TL} = {u_\rho \ov 2} \left[ \gamma_\rho \Pi_{\vp}^2 + {1 \ov \gamma_\rho}~(\partial_x \vp)^2 \right]+
\label{TL_hamiltonian_spinful} \\
&&~~~~~~+ {u_\sigma \ov 2} \left[ \gamma_\sigma \Pi_{\sigma}^2 + {1 \ov \gamma_\sigma}~(\partial_x \sigma)^2 \right],
\nonum \\
&&\mathcal{H}_C ~~= -{1 \ov m'}~\left[(\partial_x \vp)^3 +  3~\partial_x \vp~\Pi_\vp^2 +  \right.
\label{curvature_hamiltonian_spinful} \\
&&\left.~~~~~~ + 3~\partial_x \vp~[(\partial_x \sigma)^2 + \Pi_\sigma^2] + 6~\Pi_\vp~\partial_x \sigma~\Pi_\sigma  \right],
\nonum
\eea
\end{subequations}
where $m' = 6m / \sqrt{\pi}$. In Eqs.~(\ref{hamiltonian_spinful}) interactions in the charge and spin sectors have been included. For the latter we have introduced the following parameters:
\bea
\gamma_\sigma &&= \sqrt{{1 + y_{4,s}/2 - y_{2,s}/2 \ov 1 + y_{4,s}/2 + y_{2,s}/2}},
\nonum \\
u_\sigma &&= v \sqrt{(1 + y_{4,s}/2)^2 - (y_{2,s}/2)^2},
\nonum
\eea
where $y_{i,s} \equiv g_{i,s} / \pi v$ is the dimensionless coupling constant in the spin sector. Just as for the charge sector in the following we take: $y_{2,s} = y_{4,s} = y_s$, which yields:
\be
\gamma_\sigma = {1 \ov \sqrt{1 + y_s}} ~(=1), \qquad u_\sigma = v \sqrt{1 + y_s}~(=v).
\label{Spin_Parameters}
\ee
In the absence of curvature ($m' \rightarrow \infty$) only Eq.~(\ref{TL_hamiltonian_spinful}) remains which corresponds to the usual Tomonaga-Luttinger model. It's main feature is the separation of the Hilbert spaces associated with charge and spin sectors (the spin and charge parts commute with each-other; recall the anomalous commutations: $[\vp(x), \Pi_\vp(x')] = i\delta(x-x')$, and similarly for $\sigma$ and $\Pi_\sigma$, while other commutators are zero). Moreover, as can be seen from:
\bea
&&\rho_\rho(x) = -{\sqrt{2 \ov \pi}}~\partial_x \vp, \qquad \rho_\sigma(x) = -{\sqrt{2 \ov \pi}}~\partial_x \sigma,
\label{density_spinful} \\
&&~~~j_\rho(x) = {\sqrt{2 \ov \pi}}~\partial_t \vp, \qquad j_\sigma(x) = {\sqrt{2 \ov \pi}}~\partial_t \sigma,
\nonum
\eea
kinks in $\vp$ determine the density of fermions while, independently, kinks in $\sigma$ determine the magnetization density due to the spin of these fermions (we are considering the smooth part of the densities close to $q=0$; the $q=2k_F$ parts are neglected). Charge and spin travel at different velocities: $u_\rho$ and $u_\sigma$, respectively, which are determined by the interactions among the fermions. As a result, charge flies away from spin so that the interacting fermions decompose into two elementary excitations. These are the spinless charged holon and anti-holon and the chargeless spin-$1/2$ spinon, see the textbooks [\onlinecite{Book:Giamarchi,Book:GNT}]. Such elementary excitations do not appear explicitly. Only pairs of these do. They correspond to the physical excitations of the system, see also [\onlinecite{Book:Essler,Lectures:Andrei}]. In the spin sector they correspond to spin-singlet and -triplet excitations which are degenerate (they consist of two chargeless spin-$1/2$ spinons). In the charge sector they correspond to the charge-neutral holon-antiholon excitations otherwise known as particle-hole pairs. The Gaussian nature of the TL model implies that these excitations are free.

In the presence of curvature the physical excitations acquire a finite lifetime. In particular, the first two terms in Eq.~(\ref{curvature_hamiltonian_spinful}) are identical to the ones found in the spinless case. They lead to the decay of the charge excitations (particle-hole pairs) within the charge sector. More importantly, the next terms in Eq.~(\ref{curvature_hamiltonian_spinful}) show that spin and charge Hilbert spaces are no more decoupled when curvature is introduced, see Ref.~[\onlinecite{BMN}]. This opens new channels for the decay of the physical excitations. Indeed, the third and fourth terms couple the charge density to the spin density and current, respectively. And the last term describes the coupling between the charge and spin currents. Moreover, we see from Eq.~(\ref{curvature_hamiltonian_spinful}) that spin and charge degrees of freedom do not enter in a symmetric way. As a matter of fact, the decay of spin excitations can only take place by affecting the whole charge density along the wire. These field theory arguments are in qualitative agreement with results obtained from more microscopic approaches. As we know from the exact solution of the Hubbard model, {\it e.g.} Refs.~[\onlinecite{Book:Essler,Lectures:Andrei}] for reviews, the spin-charge decoupling exists at the level of the elementary excitations~\cite{Note:fractional}: holons and spinons. However,  below half-filling, physical spin excitations involve a rearrangement of charge degrees of freedom and vice versa~\cite{Note:BA}. Only in the low-energy limit, {\it i.e.} in the limit of a linear single-particle spectrum, do spin and charge sectors decouple completely. 

Even though we have derived Eq.~(\ref{hamiltonian_spinful}) the general form of the Hamiltonian and the particular asymmetry between spin and charge may be understood from general symmetry considerations. This allows a phenomenological approach to the model. As in the spinless case, the fact that all curvature terms are cubic is related to the energy of the Fermi sea for a quadratic single-particle spectrum. It implies that the particle-hole symmetry is violated which is natural away from half-filling for such a spectrum. Moreover, the curvature terms in Eq.~(\ref{curvature_hamiltonian_spinful}), are compatible with the basic symmetries of the problem: $\vp$ and $\sigma$ are odd functions of $x$ and even functions of $t$, see Eqs.~(\ref{density_spinful}). Finally, Eq.~(\ref{curvature_hamiltonian_spinful}) is invariant under $U(1)$ transformations of the charge fields and $SU(2)$ transformations of the spin fields. The latter is not quite explicit here since we are using the Abelian bosonization technique~\cite{Note:NAB}. Nevertheless, as a weak manifestation of the $SU(2)_1$ symmetry, all spin-fields in Eq.~(\ref{curvature_hamiltonian_spinful}) are paired, {\it e.g.} a term of the form $\partial_x \vp \Pi_\vp \Pi_\sigma$ is compatible with the basic symmetries but is not $SU(2)$ invariant. Another consequence of the $SU(2)$ spin-rotational invariance of the theory, {\it e.g.} see [\onlinecite{Book:GNT}], is that $y_s = -y_1/2$ where $y_1$ is the dimensionless coupling constant arising from backscattering. Thus, from Eq.~(\ref{Spin_Parameters}) and in the absence of backscattering~\cite{Note:parameters}, the spin parameters are the free ones: $u_\sigma = v$ and $\gamma_\sigma = 1$. We will work with general values of these parameters in what follows and set them to their non-interacting value at the end of the calculations.

\begin{figure}
\includegraphics[width=10cm,height=6cm]{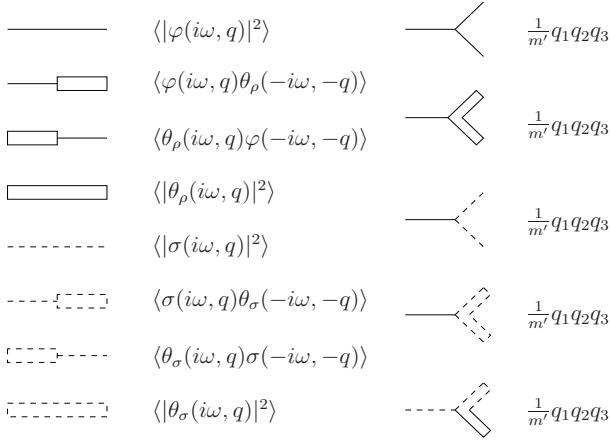}
\caption{ \label{spinful_QFT} The quantum field theory associated with the Lagrangian of Eq.~(\ref{lagrangian_spinful}) for spinful fermions.}
\end{figure}
%

\subsection{Field Theory}

To proceed further, we go to the Lagrangian formulation by introducing the dual fields: $\theta_\rho$ and $\theta_\sigma$, such that: $\Pi_\vp = \partial_x \theta_\rho$ and $\Pi_\sigma = \partial_x \theta_\sigma$. This yields:
\bea
&&\mathcal{L}_{E}[\vp,\sigma;\theta_\rho,\theta_\sigma] =
\nonum \\
&&~~~{u_\rho \gamma_\rho \ov 2}~(\partial_x \theta_\rho)^2 + {u_\rho \ov 2 \gamma_\rho}~(\partial_x \vp)^2 -i \partial_x \theta_\rho \partial_\tau \vp
\nonum \\
&&+~  {u_\sigma \gamma_\sigma \ov 2}~(\partial_x \theta_\sigma)^2 + {u_\sigma \ov 2 \gamma_\sigma}~(\partial_x \sigma)^2 -i \partial_x \theta_\sigma \partial_\tau \sigma
\nonum \\
&&-~ {1 \ov m'}~\partial_x \vp~\left[ (\partial_x \vp)^2  + 3(\partial_x \theta_\rho)^2 \right]
\nonum \\
&&-~ {3 \ov m'}~\partial_x \vp~\left[ (\partial_x \sigma)^2  + (\partial_x \theta_\sigma)^2 \right]
\nonum \\
&&-~ {6 \ov m'}~\partial_x \theta_\rho~\partial_x \sigma~\partial_x \theta_\sigma.
\label{lagrangian_spinful}
\eea
The field theory associated with Eq.~(\ref{lagrangian_spinful}) is defined on Fig.\ref{spinful_QFT}. In the charge sector, bare correlators are given by Eqs.~(\ref{bare_GF_spinless}) with $\theta \equiv \theta_\rho$. Following the spinless case the correlator $\mathcal{D}^{(0)}_{\vp \vp}$ is associated with the particle-hole pair excitations. These are {\it charge-neutral} excitations, even though we will often refer to them as charge excitations, for simplicity, as holons and anti-holons never appear explicitly. Moreover, as there is no spin-charge coupling at the level of the bare propagators, these charge-neutral excitations are spinless. In the spin sector, the bare correlators read:
\bea
&&\mathcal{D}^{(0)}_{\sigma \sigma}(i\om,q) = \la |\sigma(i\om,q)|^2 \ra = - { u_\sigma \gamma_\sigma \ov (i\om)^2 - (u_\sigma q)^2},
\nonum \\
&& \mathcal{D}^{(0)}_{\theta_\sigma \theta_\sigma}(i\om,q) = \la |\theta_\sigma(i\om,q)|^2 \ra = -{u_\sigma \ov \gamma_\sigma}~{1 \ov (i\om)^2 - (u_\sigma q)^2},
\nonum \\
&&\mathcal{D}^{(0)}_{\sigma \theta_\sigma}(i\om,q) = \la \sigma(i\om,q) \theta_\sigma(-i\om,-q) \ra =  {- i\om \ov q[(i\om)^2 -(u_\sigma q)^2]},
\nonum \\
&&\mathcal{D}^{(0)}_{\theta_\sigma \sigma}(i\om,q) = \mathcal{D}^{(0)}_{\sigma \theta_\sigma}(i\om,q),
\eea
where $\mathcal{D}^{(0)}_{\sigma \sigma}$ is associated with spin-singlet or triplet excitations. We shall often refer to them as spin excitations, for simplicity, as spinons never appear explicitly. They are chargeless.

\begin{figure}
\includegraphics[width=8.5cm,height=3cm]{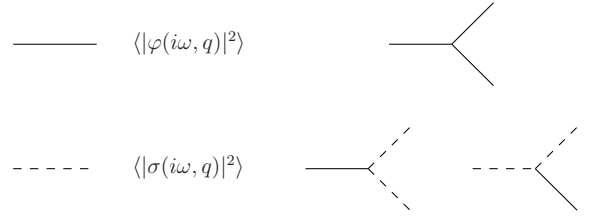}
\caption{ \label{QFT_FS} The quantum field theory related to the Lagrangian of Eq.~(\ref{lagrangian_spinful_reduced}) for spinful fermions. }
\end{figure}

Alternatively, we may also integrate over the dual fields in Eq.~(\ref{lagrangian_spinful}) and work with the following reduced Lagrangian:
\bea
\mathcal{L}_{E}[\vp,\sigma] &&= {1 \ov 2 \gamma_\rho}~\left[ {1 \ov u_\rho}~(\partial_\tau \vp)^2 + u_\rho~(\partial_x \vp)^2 \right]
\nonum \\
&&+~ {1 \ov 2 \gamma_\sigma}~\left[ {1 \ov u_\sigma}~(\partial_\tau \sigma)^2 + u_\sigma~(\partial_x \sigma)^2 \right]
\nonum \\
&&+~ {1 \ov m'v^2}~\partial_x \vp~\left[ 3 (\partial_\tau \vp)^2 - v^2 (\partial_x \vp)^2 \right]
\nonum \\
&&+~ {3 \ov m'v^2}~\partial_x \vp~\left[ (\partial_\tau \sigma)^2 - v^2 (\partial_x \sigma)^2 \right]
\nonum \\
&&+~ {6 \ov m' v^2}~\partial_\tau \vp~\partial_x \sigma~\partial_\tau \sigma,
\label{lagrangian_spinful_reduced}
\eea
where only the lowest order terms in $1/m'$ have been taken into account. The corresponding field theory is given on Fig.~\ref{QFT_FS}.

With the help of the models of Eqs.~(\ref{lagrangian_spinful}) and (\ref{lagrangian_spinful_reduced}), we will determine the effects of the spin-charge mixing terms arising from band-curvature on the response functions of the system at small wave-vector $q$. There are two of them: the charge DSF and the spin (or magnetic) DSF (mixed correlators of the type: $\la \vp(i\om,q) \sigma(-i\om,-q) \ra$, are zero within the present field theory because there is no spin-charge mixing at the level of the bare propagators).

\subsection{Dynamical charge structure factor}

\begin{figure}
\includegraphics[width=6cm,height=1cm]{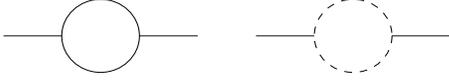}
\caption{ \label{SF_CS_PS} Second-order diagrams contributing to the charge density correlation function generated from the $\vp,\sigma-$Lagrangian of Eq.~(\ref{lagrangian_spinful_reduced}) for spinful fermions. }
\end{figure}

We first compute the second-order corrections in band-curvature to the charge DSF of spinful fermions. As for the case of spinless fermions, it corresponds to the imaginary part of the related correction for the polarization operator:
\be
\Pi_\rho^{(2)}(i\omega,q)= - {2 \gamma_\rho^2 \ov \pi}~{(u_\rho q)^2 \ov [(i \omega)^2 -(u_\rho q)^2]^2}~\Sigma_\rho(i\om,q),
\label{dcsf}
\ee
where the self-energy part corresponds to the diagrams of Fig.~\ref{SF_CS_PS} in the $\vp,\sigma-$representation. The latter include a contribution which is similar to the spinless case, the first diagram in Fig.~\ref{SF_CS_PS}, arising from the decay of a charge (-neutral) boson into two charge (-neutral) bosons, see the first vertex in Fig.~\ref{QFT_FS}. Because of band curvature, spin-charge separation is violated and spin degrees of freedom provide another channel for the decay of the charge boson. This channel corresponds to the second vertex of Fig.~\ref{QFT_FS} whereby the charge boson decays into two spin bosons (each corresponding to a spin-singlet excitation). The corresponding, second-order in band-curvature, diagram is the second diagram of Fig.~\ref{SF_CS_PS}.

\begin{figure}
\includegraphics[width=8.3cm,height=8cm]{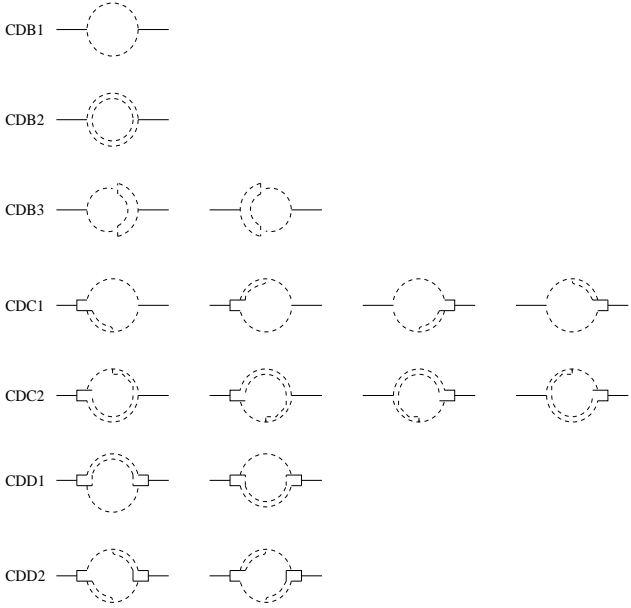}
\caption{ \label{SF_CS} Second-order diagrams mixing spin and charge degrees of freedom and contributing to the charge density correlation function of spinful fermions. These diagrams are generated with the $\vp,\sigma,\theta_\rho,\theta_\sigma-$Lagrangian of Eq.~(\ref{lagrangian_spinful}). The self-energy parts only depend on the spin degrees of freedom. CD stands for Charge Density. The set CDA is identical to the diagrams appearing in the spinless case, Fig.~\ref{Diagrams}, and is not displayed. Notice that, up to the fact that the self-energy depends only on the spin degrees of freedom, these diagrams are the same as those of the spinless case, Fig.~\ref{Diagrams}.}
\end{figure}

In the $\vp,\sigma,\theta_\rho,\theta_\sigma-$representation of Eq.~(\ref{lagrangian_spinful}) the new diagram is given by the diagrams of Fig.~\ref{SF_CS} where the same classification as in the spinless case, Fig.~\ref{Diagrams}, has been used, {\it i.e.} diagrams are classified according to the number of $\theta-$wings attached to their self-energy parts. The increased number of diagrams in Fig.~\ref{SF_CS} with respect to Fig.~\ref{SF_CS_PS} reflects the non-trivial nature of the vertices of the cubic field theory under consideration.

The diagrams of Figs.~\ref{SF_CS_PS} and \ref{SF_CS} have the same structure as those of the corresponding spinless case, Figs.~\ref{Diagrams_vp_2} and \ref{Diagrams}, respectively. In the case where charge and spin parameters are equal ($\gamma_\rho = \gamma_\sigma = \gamma$ and $u_\rho = u_\sigma = u$), the free-fermion case being a special sub-case ($\gamma = 1$ and $u=v$), the diagrams of Fig.~\ref{SF_CS_PS} reduce to the one of Fig.~\ref{Diagrams_vp_2} with an additional factor of $2$, the spin degeneracy. Similarly, the diagrams of Fig.~\ref{SF_CS} add with those of Fig.~\ref{Diagrams} and yield an overall factor of $2$. To appreciate the spin-charge coupling one has therefore to impose an inequivalence between these degrees of freedom.

This arises naturally for arbitrary values of the parameters, $\gamma_\rho \not= \gamma_\sigma$. The total self-energy, close to $q=0$, of interacting spinful fermions then reads (at $T=0$):
\bea
&&\Im \Sigma_\rho^R(\om,q) = { \pi \ov 96 }~\left( q^2 \ov m \right)^2
\label{spinful_SC} \\
&&\left[~{ \gamma_\rho^2 \ov u_\rho}~\left[ 1 + {3 \ov \gamma_\rho^2} \right]^2~u_\rho q~\left[ \delta[\om - u_\rho q] - \delta[\om + u_\rho q] \right] \right.
\nonum \\
&&\left.+{\gamma_\sigma^2 \ov u_\sigma}~\left[ 1 + {1 \ov \gamma_\sigma^2} + {2 \ov \gamma_\sigma \gamma_\rho} \right]^2 u_\sigma q \left[ \delta[\om - u_\sigma q] - \delta[\om + u_\sigma q] \right] \right]
\nonum \\
&&+ {\gamma_\rho^2 \pi \ov 64 u_\rho}~\left[ 1 - {1 \ov \gamma_\rho^2} \right]^2~\left( q \ov m u_\rho \right)^2~\left[ \om^2 - (u_\rho q)^2 \right]~\mathcal{F}_\rho(\om,q)
\nonum \\
&&+ {\gamma_\sigma^2 \pi \ov 64 u_\sigma}~\left[ 1 - {1 \ov \gamma_\sigma^2} \right]^2~\left( q \ov m u_\sigma \right)^2~\left[ \om^2 - (u_\sigma q)^2 \right]~\mathcal{F}_\sigma(\om,q),
\nonum
\eea
where the kinematic factor reads:
\be
\mathcal{F}_\sigma(\om,q) = \theta[\om - u_\sigma q] - \theta[-\om - u_\sigma q].
\label{F_sigma}
\ee
This general result shows that the density correlation function of spinful fermions has two peaks: a charge peak at $\om = u_\rho q$ but also a spin peak at $\om = u_\sigma q$ so that part of the charge spectral weight is carried by the spin-singlet. The additional spin peak to the DSF is a witness of the spin-charge coupling due to band-curvature. It arises from the opening of a new channel for the decay of charge (-neutral) excitations, {\it i.e.} into two gapless spin-singlets traveling at $u_\sigma < u_\rho$. The sharpness of this spin peak translates the coherence of the rearrangement of the background charge due to the spin-singlet. 

In the absence of backscattering, {\it i.e.} for $\gamma_\sigma = 1$ and $u_\sigma = v$,  Eq.~(\ref{spinful_SC}) simplifies as:
\bea
&&\Im \Sigma_\rho^R(\om,q) =
\label{SE_SF_SC1}\\
&&~{ \gamma_\rho^2 \pi \ov 96 u_\rho}~\left[ 1 + {3 \ov \gamma_\rho^2} \right]^2~\left( q^2 \ov m \right)^2~u_\rho q~\left[ \delta[\om - u_\rho q] - \delta[\om + u_\rho q] \right]
\nonum \\
&&+{\pi \ov 24 v}~\left[ 1 + {1 \ov \gamma_\rho} \right]^2~\left( q^2 \ov m \right)^2~v q~\left[ \delta[\om - v q] - \delta[\om + v q] \right]
\nonum \\
&&+ {\gamma_\rho^2 \pi \ov 64 u_\rho}~\left[ 1 - {1 \ov \gamma_\rho^2} \right]^2~\left( q \ov m u_\rho \right)^2~\left[ \om^2 - (u_\rho q)^2 \right]~\mathcal{F}_\rho(\om,q).
\nonum
\eea
%

\subsubsection{Multi-pair continuum}

Eq.~(\ref{spinful_SC}) shows that both charge and spin peaks have tails due to the interactions between the fermions composing the bosonic excitations.  For $\om \gg u_\rho |q|$ and from Eqs.~(\ref{spinful_SC}) and (\ref{dcsf}), the leading contribution to the charge DSF reads: 
\bea
&&S_\rho^{(2)}(\om,q) = 
\label{dsf_tail_sf} \\
&&~~{( 1 -  \gamma_\rho^2)^2 \ov 32 u_\rho}~\left( q^2 \ov m \right)^2~{ 1 \ov \om^2 - (u_\rho q)^2 }~\mathcal{F}_\rho(\om,q)
\nonum \\
&&+{( 1 -  \gamma_\sigma^2)^2 \ov 32 u_\sigma}~\left( q^2 \ov m \right)^2~\left( u_\rho \gamma_\rho \ov u_\sigma \gamma_\sigma \right)^2{ \om^2 - (u_\sigma q)^2 \ov [\om^2 - (u_\rho q)^2]^2 }~\mathcal{F}_\sigma(\om,q).
\nonum
\eea
Notice that, in the frequency range:  $u_\sigma |q| \ll \om \ll u_\rho |q|$, a small spectral weight is provided by the spin excitations.

In the limit where backscattering is neglected, Eq.~(\ref{SE_SF_SC1}), the spin peak at $\om = v|q|$ has no long-range tails and for strong repulsive interactions, $\gamma_\rho \ll 1$, is well separated from the charge peak at $\om = u_\rho |q| = v |q| / \gamma_\rho \gg v|q|$. 

\begin{figure}
\includegraphics[width=8cm,height=3cm]{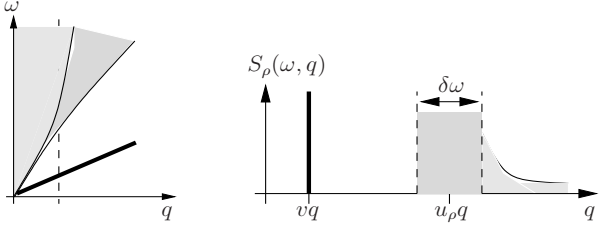}
\caption{ \label{Spectrum_SF} Schematic view on the low-momentum ($q \ll k_F$) part of the spectrum of excitation of spinful fermions and the corresponding charge DSF. Interactions are present in the charge sector ($\gamma_\rho < 1$) while the spin sector is free ($\gamma_\sigma = 1$). Because of the violation of spin-charge separation due to band-curvature  two peaks appear. The higher one corresponds to the usual charge peak and has a high-frequency tail (displayed in light gray). The lower one is the spin peak. Within bosonization, the width of the spin peak and the precise line-shape of the DSF are not accessible.}
\end{figure}
%

\subsubsection{Single-pair continuum}

We now focus on the vicinity of the charge peak,  $\om \approx u_\rho |q|$ and neglect backscattering so that charge and spin peaks are well separated from eachother. Taking into account the leading contribution, Eq.~(\ref{Pi_TL}), the total polarization operator up to second order in band-curvature reads:
\be
\Pi(i\om,q) = \Pi^{(0)}(i\om,q) + \Pi^{(2)}(i\om,q) + ...,
\ee
where the dots refer to higher order corrections and, from Eqs.~(\ref{SE_SF_SC1}) and (\ref{dcsf}),  the second order correction to the polarization operator reads:
\bea
&&\Pi^{(2)}(i\om,q) = { \gamma_\rho^2 \ov 6 \pi}~\left( q^2 \ov m \right)^2
\label{PO_SCM} \\
&&\left[~{ \gamma_\rho^2 \ov 4 u_\rho}~\left[ 1 + {3 \ov \gamma_\rho^2} \right]^2~{ (u_\rho q)^4 \ov [(i \om)^2 - (u_\rho q)^2]^3} \right.
\nonum \\
&&\left. ~+~ { 1 \ov v}~\left[ 1 + {1 \ov \gamma_\rho} \right]^2~{(v q)^2 (u_\rho q)^2 \ov [(i \om)^2 - (vq)^2][(i \om)^2 - (u_\rho q)^2]^2}~ \right].
\nonum
\eea
In Eq.~(\ref{PO_SCM}) the second term arises from the spin charge coupling. If we perform a re-summation of the second-order results, the polarization operator acquires new poles
at the frequencies solving:
\bea
&& \left [ \om^2 - (u_\rho q)^2 \right ]^2 \left [ \om^2 - (vq)^2\right ]  = 
\label{poles_sf} \\
&& ~~{2 \ov 3 \gamma_\rho}~\left( {3+ \gamma^2 \ov 4} \right)^2~\left( q^2 \ov m \right)^2~(u_\rho q)^2~[\om^2 - (v q)^2]
\nonum \\
&& + {2 \ov 3 \gamma_\rho}~\left( {1+ \gamma \ov 2} \right)^2~\left( q^2 \ov m \right)^2~(v q)^2~[\om^2 - (u_\rho q)^2].
\nonum
\eea
This equation is second order in $\om^2 - (u_\rho q)^2$ from which we recover the $2-$parametric nature of the charge peak. For $q / k_F \ll 1$, the latter reads:
\be
\om (q) \approx u_\rho^* |q| \pm {q^2 \ov 2 m^*} +  \mathcal{O}(({q  / k_F})^3),
\ee
where $m^*$ is still given by Eq.~(\ref{effective_m}). At this order the presence of the spin peak does not affect the width of the charge peak. The main effect of the spin-charge coupling, within the present re-summation based on the second order self-energy part, is then to shift the velocity of the charge excitations: $u_\rho^* \approx u_\rho [ 1 - ( q / k_F )^2]$.

On the other hand, Eq.~(\ref{poles_sf}) is first order in $\om^2 - (vq)^2$. This implies that, within a re-summation based on the second order self-energy part, the
spin peak remains single-parametric with a velocity slightly renormalized by the band-curvature: $v^* \approx v [ 1 - ( q / k_F )^2]$.

Our results are schematically displayed on Fig.~\ref{Spectrum_SF}. Within the present, perturbative, approach we cannot access the precise line-shape of the charge DSF in the vicinity of $\om \approx u_\rho |q|$ and $\om \approx v |q|$. 

\subsection{Dynamical spin structure factor}

We now focus on the spin degrees of freedom. The long-distance part (close to $q=0$; the $2k_F-$part is neglected) of the spin-density correlation function is defined as:
\be
\chi(x,\tau) = {2 \ov \pi}~\la \partial_x \sigma(x,\tau) \partial_x \sigma(0,0) \ra,
\ee
and, in Fourier space, reads:
\be
\chi(i\om,q) = - {q^2 \ov \pi}~\mathcal{D}_{\sigma \sigma}(i\om,q).
\ee
The dynamic spin structure factor corresponds to the dissipative part of this susceptibility:
\be
S_\sigma(\om,q) = - \Im \chi^R(\om,q).
\ee
In the absence of curvature the result is well known:
\bea
S^0_\sigma(\om,q) =  {\gamma_\sigma \ov u_\sigma}~(u_\sigma q)^2~\delta[\omega - u_\sigma |q|],
\label{magnetic_dsf_0}
\eea
with obvious similarity with the dynamical charge structure factor.

\begin{figure}
\includegraphics[width=6cm,height=1cm]{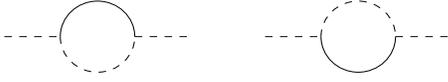}
\caption{ \label{SF_SS_PS} Second-order diagrams contributing to the spin-density correlation function and generated from the $\vp,\sigma-$Lagrangian of Eq.~(\ref{lagrangian_spinful_reduced}) for spinful fermions.}
\end{figure}

Including curvature, the second-order correction to the spin density correlation function reads:
\be
\chi^{(2)}(i\om,q) = - {2 \gamma_\sigma^2 \ov \pi}~{(u_\sigma q)^2 \ov [(i\om)^2 - (u_\sigma q)^2]^2}~\Sigma_\sigma(i \om, q),
\ee
where the self-energy $\Sigma_\sigma$ is given by the diagrams of Fig.~\ref{SF_SS_PS} in the the $\vp,\sigma-$representation of Lagrangian Eq.~(\ref{lagrangian_spinful_reduced}). These diagrams are built from the third vertex displayed on Fig.~\ref{QFT_FS} which shows that a spin excitation (singlet, triplet) can decay only into a mixed spin (singlet, triplet, resp.) $-$ charge-neutral excitation. Contrary to charge fluctuations which had two-channels through which they could decay (one of them not affecting the spin degrees of freedom and therefore surviving in the spinless case) the decay of spin fluctuations cannot proceed without affecting the whole charge background.

\begin{figure}
\includegraphics[width=8cm,height=11cm]{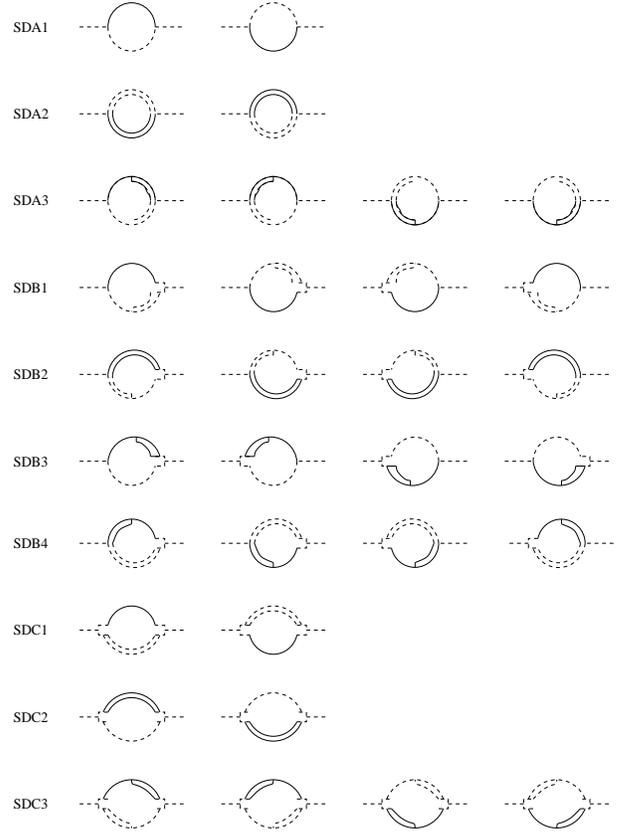}
\caption{ \label{SF_SS} Second-order diagrams contributing to the spin-density correlation function and generated from the $\vp,\sigma,\theta_\rho,\theta_\sigma-$Lagrangian of Eq.~(\ref{lagrangian_spinful}) for spinful fermions. All the self-energy parts mix spin and charge. SD stands for Spin Density. In the limit where spin and charge degrees of freedom are equal these diagrams reduce to those of Fig.~\ref{Diagrams} up to spin degeneracy.}
\end{figure}

In the $\vp,\sigma,\theta_\rho,\theta_\sigma-$representation of Lagrangian Eq.~(\ref{lagrangian_spinful}), which reveals the non-trivial nature of the vertices of the cubic field theory, these mixed spin-charge self-energy parts are displayed on Fig.~\ref{SF_SS}.

In the case where spin and charge degrees of freedom are equivalent ($\gamma_\rho = \gamma_\sigma = \gamma$ and $u_\rho = u_\sigma = u$), the free-fermion case being a special sub-case ($\gamma = 1$ and $u=v$), spin and charge bare propagators are equal so that there is no more difference between the solid and dashed lines in Figs.~\ref{SF_SS_PS} and \ref{SF_SS}. One can then easily check that in this case the diagrams of Figs.~\ref{SF_SS_PS} and \ref{SF_SS} are identical to those of Figs.~\ref{Diagrams_vp_2} and \ref{Diagrams}, respectively, with an additional factor of $2$ due to the spin degeneracy.


All non-trivial effects that we shall observe below are therefore related to the fact that spin and charge excitations travel at different velocities ($\gamma_\rho \not= \gamma_\sigma$). Even though calculations may be carried out for arbitrary spin and charge parameters we will focus here, for simplicity, on the case where backscattering is neglected so that the spin sector is free ($\gamma_\sigma = 1$ and $u_\sigma=v$). The self-energy then reads (at $T=0$ for simplicity):
\bea
&&\Im \Sigma_\sigma^R(\om,q) =
\label{SE_SF_SS2} \\
&&~~{ \gamma_\rho \pi \ov 8 v}~\left[ 1 + {1 \ov \gamma_\rho} \right]^2~\left[ 1 + {\om \ov v q} \right]^2~\left({q^2 \ov m} \right)^2~v q
\nonum \\
&&{[\om - v q]~[\om - u_\rho q] \ov |u_\rho - v|^3 q^3}~\left[ \theta \left[ -{u_\rho [\om - v q] \ov u_\rho - v}\right] - \theta \left[ -{v [\om - u_\rho q] \ov u_\rho - v}\right] \right]
\nonum \\
&&+{ \gamma_\rho \pi \ov 8 v}~\left[ 1 + {1 \ov \gamma_\rho} \right]^2~\left[ 1 - {\om \ov v q} \right]^2~\left({q^2 \ov m} \right)^2~v q
\nonum \\
&&{[\om + v q]~[\om + u_\rho q] \ov |u_\rho - v|^3 q^3}~\left[ \theta \left[ -{u_\rho [\om + v q] \ov u_\rho - v}\right] - \theta \left[ -{v [\om + u_\rho q] \ov u_\rho - v}\right] \right]
\nonum \\
&&+{ \gamma_\rho \pi \ov 8 }~\left[ 1 - {1 \ov \gamma_\rho} \right]^2~\left[ 1 - {\om \ov v q} \right]^2~\left({q \ov m} \right)^2~{1 \ov |u_\rho + v|^3}
\nonum \\
&&[\om + v q]~[\om - u_\rho q]~\left[ \theta \left[ {v [\om - u_\rho q] \ov u_\rho + v}\right] - \theta \left[ -{u_\rho [\om + v q] \ov u_\rho + v}\right] \right]
\nonum \\
&&+{ \gamma_\rho \pi \ov 8}~\left[ 1 - {1 \ov \gamma_\rho} \right]^2~\left[ 1 + {\om \ov v q} \right]^2~\left({q \ov m} \right)^2~{1 \ov |u_\rho + v|^3}
\nonum \\
&&[\om - v q]~[\om + u_\rho q]~\left[ \theta \left[ {v [\om + u_\rho q] \ov u_\rho + v}\right] - \theta \left[- {u_\rho [\om - v q] \ov u_\rho + v}\right] \right].
\nonum
\eea
Notice that in the limit where $\gamma_\rho \rightarrow 1$ in Eq.~(\ref{SE_SF_SS2}) the first two terms become $\delta-$functions centered around $\pm vq$ and the last two terms vanish. On the other hand for $\gamma_\rho < 1$, the combined effect of repulsive interactions and second-order curvature effects broaden the peaks at $\pm vq$ by transferring spectral weight to frequencies reaching the charge frequency ($u_\rho q$) with additional long-range tails. This is more conveniently seen on the magnetic DSF. The latter is non-zero all the way between the spin and charge frequencies:
\bea
&&S_\sigma^{(2)}(\om,q) =  { \gamma_\rho \ov 4 v}~\left({q \ov m v} \right)^2~\left[ \theta [ - \om + u_\rho q ] - \theta[ - \om + v q] \right]
\nonum \\
&&\left[ {[ 1 / \gamma_\rho +1 ]^2 \ov |1 / \gamma_\rho - 1|^3 }~{\om - u_\rho q \ov \om - vq } + {[ 1 / \gamma_\rho - 1 ]^2 \ov |1/\gamma_\rho + 1|^3 }~{\om + u_\rho q \ov \om - v q} \right],
\label{sdsf}
\eea
where, for simplicity, we have set $q>0$ and focused on $\om > vq$.

At $\om = vq$ this expression diverges but only algebraically and not as a $\delta-$function as for the zero-order term or in the absence of interactions.

At $\om = u_\rho q$ the magnetic DSF is finite and equals:
\be
S_\sigma^{(2)}(u_\rho q,q) = {1  \ov 2 v}~{ 1 / \gamma_\rho - 1 \ov |1/\gamma_\rho + 1|^3 }~\left({q \ov m v} \right)^2,
\label{sdsf_holon}
\ee
so that part of the magnetic spectral weight is carried by the charge excitations.

At large frequencies, $\om \gg u_\rho q$, tails appear and read:
\bea
S_\sigma^{(2)}( \om ,q) &&= { \gamma_\rho \ov 2 v}~\left[{ 1 / \gamma_\rho - 1  \ov 1/\gamma_\rho + 1 }\right]^2~\left({q^2 \ov m} \right)^2~{\theta[\om - u_\rho q] \ov \om^2 - (vq)^2}
\nonum \\
&&+~{ \gamma_\rho \ov 2 v}~{ [ 1 / \gamma_\rho - 1 ]^2 \ov |1/\gamma_\rho + 1|^3 }~\left({q \ov m v} \right)^2~\theta[\om - u_\rho q].
\label{sdsf_tails}
\eea
Notice that the second term in Eq.~(\ref{sdsf_tails}) yields a non-zero dissipation at infinite frequencies. Nevertheless, the spin-structure factor satisfies the sum-rule:
\bea
\int_0^{\infty}~{{d \om}}~\om~S_\sigma(\om ,q) = {v^* q^2 \ov 2},
\label{sum_rule_sdsf}
\eea
but with a re-normalized spin-velocity, $v^*$, which reads:
\be
{v^* \ov v} = 1 + {\gamma_\rho^2 \ov 4}~{ [ 1 - \gamma_\rho ]^2 \ov |1 + \gamma_\rho |^3}, \qquad v < v^* < u_\rho.
\label{spin_dressed}
\ee
In Eq.~(\ref{sum_rule_sdsf}) a large frequency cut-off at $\eps_F = m v^2$ has been included and the final result was obtained after sending this cut-off to infinity. The re-normalization of the spin-velocity by interactions in the charge sector is precisely due to the last term in Eq.~(\ref{sdsf_tails}). It is a signature that the physical excitations carrying the magnetic spectral weight, in the tails of the spin-structure factor, are spin (-singlet or -triplet) excitations dressed by charge-neutral ones.

\begin{figure}
\includegraphics[width=8cm,height=5cm]{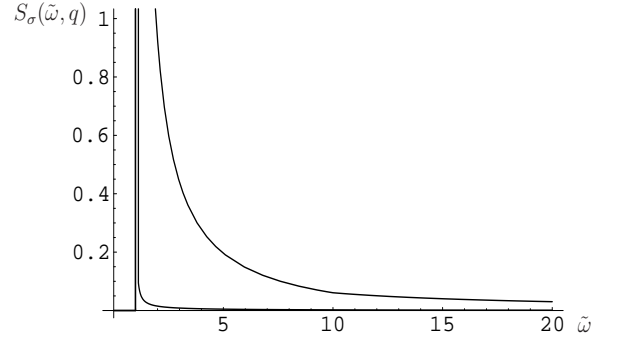}
\caption{ \label{SS_Full} The magnetic structure factor, $S_\sigma({\tilde \om},q)$, at a fixed $q$, as a function of the dimensionless frequency ${\tilde \om} = \om / vq$ and for different interaction strengths in the charge sector ($\gamma_\sigma = 1$). This figure shows a transfer of the spectral weight to higher frequencies as the interaction strength increases. This transfer originates from the spin-charge coupling due to band-curvature. The spectral weight then extends from the spin excitation frequency, ${\tilde \om}_\sigma = 1$, to the interaction-dependent charge excitation frequency, ${\tilde \om}_\rho = 1/\gamma_\rho$, with additional high frequency tails. The peak at ${\tilde \om} = 1$ corresponds to the free fermion result: $\gamma_\rho = 1$ (${\tilde \om}_\sigma = {\tilde \om}_\rho = 1$). The second curve corresponds to the second-order curvature correction at $\gamma_\rho=0.9$ (${\tilde \om}_\rho = 1.1$). The third curve corresponds to the second-order curvature correction at $\gamma_\rho=0.1$ (${\tilde \om}_\rho = 10$).}
\end{figure}

We therefore see that band-curvature softens the singularities of the spin structure factor already at the second order of perturbation theory. Such an effect arises from spin-charge mixing and is more dramatic than for the charge DSF. The latter was also affected by the violation of the spin-charge separation due to band curvature ({\it cf.} the double peak structure, see Eqs.~(\ref{SE_SF_SC1}) and~(\ref{PO_SCM}) and Fig.~\ref{Spectrum_SF}) but its low frequency part was still $\delta-$singular at second order in band-curvature. The sharpness of both charge and spin peaks in the charge DSF was
due to the decay of charge fluctuations into either a pair of charge-neutral excitations or a pair of spin-singlet excitations, both coherently traveling at velocities $u_\rho$ and $u_\sigma$, respectively. On the other hand spin fluctuations decay via spin(-singlet or -triplet) $-$  charge(-neutral) excitations. Both components of this pair lead to rearrangement of the background charge along the wire~\cite{Note:BA}. But they travel at different velocities. As a consequence, the redistribution of magnetic spectral weight is incoherent and affects an infinite number of modes. The magnetic DSF is plotted on Fig.~\ref{SS_Full} for different values of the interactions in the charge sector. Because of the spin-charge coupling just described more spectral weight is transfered from the spin peak at $\om = vq$ to the charge peak at $\om = u_\rho q = v q / \gamma_\rho$, upon increasing interactions.

\section{Conclusion}
\label{Conclusion}

Using the bosonization technique we have studied the charge and spin equilibrium dynamics of 1D fermions in clean quantum wires with forward scattering interactions, {\it i.e.} below half-filling, and band-curvature, {\it i.e.} quadratic, corrections to the linear single-particle spectrum, {\it cf.} Eq.~(\ref{non_linear_spectrum}). The dynamics were accessed by calculating the charge and magnetic dynamical structure factors at small wave-vector $q$. Charge dynamics may be accessed experimentally with the help of energy and momentum resolved spectroscopies, [\onlinecite{Auslender}], electronic Raman spectroscopy, [\onlinecite{Raman}], or electron energy loss spectroscopies [\onlinecite{EEL}].

The introduction of band-curvature corrections allowed us to explore the high-energy physics of fermions in quantum wires. This a basic step beyond the standard Tomonaga-Luttinger model which assumes a linear dispersion, Eq.~(\ref{spectrum_Luttinger}). In solid-state physics this relativistic approximation is valid only at low-energies. It is also a basic step towards the Hubbard model (below half-filling) which has the full non-linear spectrum of lattice fermions, Eq.~(\ref{spectrum_Hubbard}).

Based on recent results for fermions without spin [\onlinecite{Glazman:DSF_2006,Affleck}] we have discussed the efficiency and limitations of bosonization in presence of curvature corrections,  {\it cf.} Sec.~\ref{SL_section}. Our main result is Eq.~(\ref{SE_SF}) for the self-energy part of the DSF, in second-order in band-curvature and to all orders in interactions. From this self-energy the $2-$pair continuum of the DSF has been derived, following Ref.~[\onlinecite{Affleck}], and in agreement with other known results, {\it cf.}~[\onlinecite{Glazman:DSF_2003,ST}]. Moreover, a re-summation of band-curvature effects based on the second-order self-energy enabled us to emulate the 2-parametric structure of the single-pair continuum, cf. Eq.~(\ref{effective_m}), in qualitative agreement with exact results [\onlinecite{Glazman:DSF_2006,Affleck}]. On the other hand the precise line shape of the DSF could not be derived with the help of this self-energy and a more efficient re-summation scheme has to be developed.

In the case of fermions with spin we have focused on the spin-charge coupling which arises from the interplay between curvature and interactions, {\it cf.} Sec.~\ref{SF_section}, following Ref.~[\onlinecite{BMN}]. The spin-charge coupling manifests by the opening of new channels for the decay of the excitations. For charge excitations, besides the usual decay into two other charge-neutral excitations, as in the spinless case, a decay into two spin-singlets takes place. On the other hand there is a unique channel of decay of spin (-singlet or triplet) excitations: via a pair composed of the spin excitation itself and a charge-neutral excitation. The decay of spin excitations therefore always affects the charge degrees of freedom.

For the charge dynamics our main result is Eq.~(\ref{spinful_SC}) the self-energy part of the charge DSF, in second-order in band-curvature and to all orders in interactions. This equation and the corresponding Fig.~\ref{Spectrum_SF} show that the charge DSF has a double peak structure: a charge peak at $\om = u_\rho |q|$ but also a spin peak at $\om = v|q|$. The charge peak corresponds to an incoherent continuum of single spinless charge-neutral excitations. It has a high-frequency tail corresponding to the incoherent continuum of pairs of these spinless charge-neutral excitations. On the other hand the spin peak corresponds to the incoherent continuum of chargeless single spin-singlet excitations. A re-summation based on the second order self-energy has revealed that the 2-parametric structure of the charge peak is qualitatively similar to the case of spinless fermions. The width of the spin peak as well as the precise line-shape of the charge DSF could not be derived with the help of this self-energy. For the spin dynamics our main result is Eq.~(\ref{SE_SF_SS2}), the self-energy part of the spin DSF, in second-order in band-curvature and to all orders in interactions. This self-energy displays a transfer of magnetic spectral weight to higher frequencies. This continuous transfer extends from the spin frequency $\om = v |q|$ to the charge frequency $\om = u_\rho |q|$ with additional tails at $\om \gg u_\rho |q|$. Because, $u_\rho = v / \gamma_\rho$ is interaction-dependent the transfer of magnetic spectral weight to higher frequencies increases with increasing repulsive interactions, see Fig.~\ref{SS_Full}.

\acknowledgments

I thank the Abdus Salam ICTP for hospitality. I am deeply grateful to S. Brazovskii, F. H. L. Essler and A. A. Nersesyan for fruitful discussions as well as the referees for their
criticism and suggestions.

\appendix

\section{Notations}
\label{A:Notations}

Our notations are different from the ones often found in the literature, {\it cf.} [\onlinecite{Haldane,Samokhin,Affleck}]. They have:
\be
\mathcal{H}_{Curvature} = {1 \ov 12 \pi m} \left[ :(\partial_x \vp_R(x))^3: - :(\partial_x \vp_L(x))^3: \right],
\nonum
\ee
with: $\psi_{R(L)}(x) \sim e^{\pm ik_Fx - i \vp_{R(L)}}$. We have therefore the following correspondence between these notations and our notations:
\be
\vp_+ \rightarrow - \vp_R / \sqrt{4 \pi}, \qquad \vp_- \rightarrow  \vp_L / \sqrt{4 \pi}.
\label{other_notation}
\ee
We may check that, indeed, this yields the same numerical factor: $(2 \sqrt{\pi} /3m)(1 / 8 \pi \sqrt{\pi}) = 1 / 12 \pi m$.

Moreover, the chiral basis, $\vp_\pm$, is often used. We which to relate the general irrelevant operator expressed in the $\vp$ basis:
\be
\partial_x \vp \left[ n(\partial_\tau \vp)^2 - v^2 (\partial_x \vp)^2 \right],
\nonum
\ee
with its expression in the chiral basis ($n = 3 / \gamma_\rho^2$ in the text). In real time this can be done with the help of: $(1/v) \partial_t \vp = \Pi_\vp = -\partial_x \vp_+ + \partial_x \vp_-$ and $\partial_x \vp = \partial_x \vp_+ + \partial_x \vp_-$, and yields:
\bea
&&\partial_x \vp \left[ n(\partial_\tau \vp)^2 - v^2 (\partial_x \vp)^2 \right] =
\nonum \\
&&-(n+1)~\left[ (\partial_x \vp_+)^3 + (\partial_x \vp_-)^3 \right] +
\nonum \\
&&+(n-3)~\left[(\partial_x \vp_+)^2~\partial_x \vp_- + \partial_x \vp_+~(\partial_x \vp_-)^2 \right].
\nonum
\eea
The case $n=3$ corresponds to free fermions:
\bea
\partial_x \vp \left[ 3(\partial_\tau \vp)^2 - v^2 (\partial_x \vp)^2 \right] &&= -4~\left[ (\partial_x \vp_+)^3 + (\partial_x \vp_-)^3 \right]
\nonum \\
&&= -{1 \ov \pi}~\left[ (\partial_x \vp_R)^3 + (\partial_x \vp_L)^3 \right],
\nonum
\eea
and is the sum of two chiral terms. This implies that tails are in the crossed chiral terms. They can be isolated by a special (unrealistic) choice of the interactions: $n=-1$, which yields:
\bea
&&\partial_x \vp \left[ (\partial_\tau \vp)^2 + v^2 (\partial_x \vp)^2 \right] =
\nonum \\
&&4 \left[(\partial_x \vp_+)^2~\partial_x \vp_- + \partial_x \vp_+~(\partial_x \vp_-)^2 \right] =
\nonum \\
&&{1 \ov \pi \sqrt{4 \pi}} \left[(\partial_x \vp_R)^2~\partial_x \vp_L - \partial_x \vp_R~(\partial_x \vp_L)^2 \right],
\nonum
\eea
where Eq.~(\ref{other_notation}) has been used.


\section{Operator product expansions}
\label{A:OPE}

Bosonization rests on the following vertex operator:
\be
\psi_\pm(x) = {1 \ov \sqrt{2 \pi}}~:\exp \left[ \pm i \sqrt{4 \pi}~\vp_\pm(x) \right]:,
\nonum
\ee
relating the chiral fermionic fields to the chiral bosonic fields. One then needs to express various operators (density, Hamiltonian...) made out of fermions in terms of the bosonic fields. The operators contain the products of fermionic operators at coinciding points and a regularization procedure is required in order to remove the short-distance divergences. This can be achieved by point-splitting the operators, {\it {\it e.g.}}:
\bea
\psi_\eta^\dagger(x) \psi_\eta(x) = {\mathrm{lim}}_{\eps \rightarrow 0} \left[\psi_\eta^\dagger(x+\eps)  \psi_\eta(x) - \la \psi_\eta^\dagger(x+\eps)  \psi_\eta(x) \ra \right],
\nonum
\eea
where $\eta = \pm$ and:
\be
\la \psi_\pm^\dagger(x+\eps)  \psi_\pm(x) \ra = \pm {1 \ov 2 \pi i \eps}.
\label{Fermi_correlations}
\ee
The product of vertex operators is normal ordered according to:
\be
:e^{i\beta \vp_\eta(x)}::e^{i\beta' \vp_\eta(x')}: = :e^{i\beta \vp_\eta(x)+ i\beta' \vp_\eta(x')}: e^{-\beta \beta' \la \vp_\eta(x) \vp_\eta(x') \ra}
\nonum
\ee
where:
\be
\la \vp_\eta(x) \vp_\eta(x') \ra = -{1 \ov 4 \pi}~\ln[i(x-x')].
\label{Bose_correlations}
\ee
Notice that the average values in Eqs.~(\ref{Fermi_correlations}) and (\ref{Bose_correlations}) were taken over the TL action, {\it i.e.} no curvature correction was taken into account. We believe this approximation is sufficient, within the present perturbative approach, to bosonize the theory. Indeed, the first curvature corrections to, {\it e.g.} Eq.~(\ref{Bose_correlations}), are in $1/m^2$ while the fermionic Hamiltonian has a contribution in $1/m$. At the level of the current algebra this approximation implies that we keep the usual (ultra-local) commutation relations:
\be
[\partial_x \vp(x), \Pi(x')] = i \partial_x \delta(x-x') + ...,
\nonum
\ee
where the dots correspond to the neglected curvature corrections.

Altogether this yields:
\bea
\rho_\eta(x) &&= \psi_\eta^\dagger(x) \psi_\eta(x)
\nonum \\
&&= {\mathrm{sgn}}(\eta)~{\mathrm{lim}}_{\eps \rightarrow 0}~{1 \ov 2 \pi i \eps}~\left[ e^{-i {\mathrm{sgn}}(\eta) \sqrt{4 \pi} \eps \partial_x \vp_\eta(x)} - 1 \right]
\nonum
\eea
The first OPE reads:
\bea
\rho_\eta(x) = \psi_\eta^\dagger(x) \psi_\eta(x) = - {1 \ov \sqrt{\pi}}~\partial_x \vp_\eta(x).
\eea
More generally:
\bea
&&\psi_\eta^\dagger(y) \psi_\eta(x) =
\nonum \\
&&{\mathrm{sgn}}(\eta)~{1 \ov 2 \pi i (y - x)}~\left[ e^{-i {\mathrm{sgn}}(\eta) \sqrt{4 \pi} [\vp_\eta(y) - \vp_\eta(x)]} - 1 \right],
\nonum
\eea
from which:
\bea
&&\psi_\eta^\dagger(y) \partial_x \psi_\eta(x) =
\nonum \\
&&{\mathrm{sgn}}(\eta)~{1 \ov 2 \pi i (y - x)^2}~\left[ e^{-i {\mathrm{sgn}}(\eta) \sqrt{4 \pi} [\vp_\eta(y) - \vp_\eta(x)]} - 1 \right]
\nonum \\
&&+ {1 \ov \sqrt{\pi} (y-x)}~\partial_x \vp_\eta(x)~e^{-i {\mathrm{sgn}}(\eta) \sqrt{4 \pi} [\vp_\eta(y) - \vp_\eta(x)]},
\nonum
\eea
and:
\bea
&&\psi_\eta^\dagger(y) \partial_{xx} \psi_\eta(x) =
\nonum \\
&&{\mathrm{sgn}}(\eta)~{1 \ov \pi i (y - x)^3}~\left[ e^{-i {\mathrm{sgn}}(\eta) \sqrt{4 \pi} [\vp_\eta(y) - \vp_\eta(x)]} - 1 \right]
\nonum \\
&&+ {2 \ov \sqrt{\pi} (y-x)^2}~\partial_x \vp_\eta(x)~e^{-i {\mathrm{sgn}}(\eta) \sqrt{4 \pi} [\vp_\eta(y) - \vp_\eta(x)]}
\nonum \\
&&+ {1 \ov \sqrt{\pi} (y-x)}~\partial_{xx} \vp_\eta(x)~e^{-i {\mathrm{sgn}}(\eta) \sqrt{4 \pi} [\vp_\eta(y) - \vp_\eta(x)]}
\nonum \\
&&+ {\mathrm{sgn}}(\eta)~{2i \ov y-x}~(\partial_x \vp_\eta(x))^2~e^{-i {\mathrm{sgn}}(\eta) \sqrt{4 \pi} [\vp_\eta(y) - \vp_\eta(x)]}.
\nonum
\eea
The short-distance products are obtained by setting: $y = x + \eps$ and expanding to the lowest meaningful order in $\eps$.
The diverging terms in $1/\eps$ cancel out and as a result the second OPE reads:
\be
\psi_\eta^\dagger(x) \partial_x \psi_\eta(x) = - i~{\mathrm{sgn}}(\eta)~:(\partial_x \vp_\eta(x))^2: - {1 \ov 2 \sqrt{\pi}}~\partial_{xx} \vp_\eta(x).
\ee
It may be introduced into the Dirac Hamiltonian density:
\bea
\mathcal{H}_{Dirac} &&= iv \left[ \psi_+^\dagger(x) \partial_x \psi_+(x) - \psi_-^\dagger(x) \partial_x \psi_-(x) \right]
\nonum \\
&&= v \left[ :(\partial_x \vp_+(x))^2: + :(\partial_x \vp_-(x))^2: \right],
\nonum
\eea
where we have neglected the terms which vanish upon integrating over space.

Similarly, the curvature terms read:
\bea
&&\psi_\eta^\dagger(x+\eps) \partial_{xx} \psi_\eta(x) = {\mathrm{sgn}}(\eta)~{1 \ov \pi i \eps^3}
\nonum \\
&&\left[ e^{-i {\mathrm{sgn}}(\eta) \sqrt{4 \pi} [\eps \partial_x \vp_\eta(x) + \eps^2/2~\partial_{xx} \vp_\eta(x) + \eps^3/6~\partial_{xxx} \vp_\eta(x)]} - 1 \right] \nonum \\
&&+ {2 \ov \sqrt{\pi} \eps^2}~\partial_x \vp_\eta(x)~e^{-i {\mathrm{sgn}}(\eta) \sqrt{4 \pi} [\eps \partial_x \vp_\eta(x) + \eps^2/2~\partial_{xx} \vp_\eta(x)]}
\nonum \\
&&+ {1 \ov \sqrt{\pi} \eps}~\partial_{xx} \vp_\eta(x)~e^{-i {\mathrm{sgn}}(\eta) \sqrt{4 \pi} \eps \partial_x \vp_\eta(x)}
\nonum \\
&&+ {\mathrm{sgn}}(\eta)~{2i \ov \eps}~(\partial_x \vp_\eta(x))^2~e^{-i {\mathrm{sgn}}(\eta) \sqrt{4 \pi} \eps \partial_x \vp_\eta(x)}
\nonum
\eea
Expanding in $\eps$ the diverging terms in $1/\eps$ and $1/\eps^2$ cancel out and the third OPE reads:
\bea
&&\psi_\eta^\dagger(x) \partial_{xx} \psi_\eta(x) = {4 \sqrt{\pi} \ov 3}~:(\partial_x \vp_\eta(x))^3:
\\
&&- 3 i {\mathrm{sgn}}(\eta)~:\partial_x \vp_\eta(x)\partial_{xx} \vp_\eta(x): - {1 \ov 3 \sqrt{\pi}}~\partial_{xxx} \vp_\eta(x).
\nonum
\eea
It may be introduced into the curvature part of the Hamiltonian density:
\bea
\mathcal{H}_{Curvature} &&= -{1 \ov 2m} \left[ \psi_+^\dagger(x) \partial_{xx} \psi_+(x) + \psi_-^\dagger(x) \partial_x \psi_-(x) \right]
\nonum \\
&&= -{2 \sqrt{\pi} \ov 3m} \left[ :(\partial_x \vp_+(x))^3: + :(\partial_x \vp_-(x))^3: \right],
\nonum
\eea
where we have neglected the terms which vanish upon integrating over space or by symmetry considerations [$\vp$ is an odd function of $x$ so that $(\partial_x \vp)^3$ is even but $\partial_x \vp~\partial_{xx} \vp$ is odd and does not contribute to the energy].


\begin{widetext}

\section{Correlation functions in the spinless fermion case}

\subsection{Second-order perturbation in curvature: $\vp-$representation}
\label{A:SL1}

The second order correlation function reads:
\bea
&&\la |\vp(i\om,q)|^2 \ra^{(2)} = {1 \ov 2 (m'u_\rho^2)^2}~{1 \ov \beta^6}~\sum_{\om_1,\om_2,\om_3,\om_1',\om_2',\om_3'}~\int {{dq_1 dq_2 dq_3 \ov (2 \pi)^3} {dq_1' dq_2' dq_3' \ov (2 \pi)^3}}~\delta[1+2+3] \delta[1'+2'+3']
\nonum \\
&&\left[n_\rho~q_1 \om_2 \om_3 - u_\rho^2~q_1 q_2 q_3 \right]\left[n_\rho~q_1' \om_2' \om_3' - u_\rho^2~q_1' q_2' q_3' \right] \la  \vp_{\om}(q) \vp_{-\om}(-q) \vp_{\om_1}(q_1) \vp_{\om_2}(q_2) \vp_{\om_3}(q_3) \vp_{\om_1'}(q_1') \vp_{\om_2'}(q_2') \vp_{\om_3'}(q_3') \ra_0,
\nonum
\eea
where $n_\rho = 3/\gamma_\rho^2$ and may be conveniently split into three contributions:
\bea
&&\la |\vp(i\om,q)|^2 \ra_A^{(2)} = {u_\rho^4 \ov 2 (m'u_\rho^2)^2}~{1 \ov \beta^6}~\sum_{\om_1,\om_2,\om_3,\om_1',\om_2',\om_3'}~\int {{dq_1 dq_2 dq_3 \ov (2 \pi)^3} {dq_1' dq_2' dq_3' \ov (2 \pi)^3}}~\delta[1+2+3] \delta[1'+2'+3']
\nonum \\
&& q_1 q_2 q_3 q_1' q_2' q_3'~\la  \vp_{\om}(q) \vp_{-\om}(-q) \vp_{\om_1}(q_1) \vp_{\om_2}(q_2) \vp_{\om_3}(q_3) \vp_{\om_1'}(q_1') \vp_{\om_2'}(q_2') \vp_{\om_3'}(q_3') \ra_0,
\nonum
\eea
\bea
&&\la |\vp(i\om,q)|^2 \ra_B^{(2)} = {n_\rho^2 \ov 2 (m'u_\rho^2)^2}~{1 \ov \beta^6}~\sum_{\om_1,\om_2,\om_3,\om_1',\om_2',\om_3'}~\int {{dq_1 dq_2 dq_3 \ov (2 \pi)^3} {dq_1' dq_2' dq_3' \ov (2 \pi)^3}}~\delta[1+2+3] \delta[1'+2'+3']
\nonum \\
&& q_1 \om_2 \om_3 q_1' \om_2' \om_3'~\la  \vp_{\om}(q) \vp_{-\om}(-q) \vp_{\om_1}(q_1) \vp_{\om_2}(q_2) \vp_{\om_3}(q_3) \vp_{\om_1'}(q_1') \vp_{\om_2'}(q_2') \vp_{\om_3'}(q_3') \ra_0,
\nonum
\eea
\bea
&&\la |\vp(i\om,q)|^2 \ra_C^{(2)} = - {n_\rho u_\rho^2 \ov (m'u_\rho^2)^2}~{1 \ov \beta^6}~\sum_{\om_1,\om_2,\om_3,\om_1',\om_2',\om_3'}~\int {{dq_1 dq_2 dq_3 \ov (2 \pi)^3} {dq_1' dq_2' dq_3' \ov (2 \pi)^3}}~\delta[1+2+3] \delta[1'+2'+3']
\nonum \\
&& q_1 \om_2 \om_3 q_1' q_2' q_3' ~\la  \vp_{\om}(q) \vp_{-\om}(-q) \vp_{\om_1}(q_1) \vp_{\om_2}(q_2) \vp_{\om_3}(q_3) \vp_{\om_1'}(q_1') \vp_{\om_2'}(q_2') \vp_{\om_3'}(q_3') \ra_0.
\nonum
\eea
The first contribution is easy to Wick order and yields:
\bea
&&\la |\vp(i\om,q)|^2 \ra_A^{(2)} = {18 q^2 u_\rho^4 \ov (m'u_\rho^2)^2}~[\mathcal{D}(i\om,q)]^2~{1 \ov \beta}~\sum_{\nu}~\int {{dp \ov 2 \pi}}~p^2 (p-q)^2~\mathcal{D}(i\nu,p)\mathcal{D}(i\nu-i\om,p-q).
\nonum
\eea
The imaginary part of the corresponding retarded self-energy reads:
\bea
\Im \Sigma_A^R(\om,q) &&= {3 \ov 8 u_\rho}~\left( {v \ov u_\rho} \right)^2~\left( q^2 \ov m' \right)^2~u_\rho q~\left[ \delta[\om - u_\rho q] - \delta[\om + u_\rho q] \right]
\nonum \\
&&+ {9 \ov 16 u_\rho}~\left( {v \ov u_\rho} \right)^2~\left( q \ov m' u_\rho \right)^2~\left[ \om^2 - (u_\rho q)^2 \right]~\mathcal{F}_\rho(\om,q),
\eea
where $\mathcal{F}_\rho$ is given by Eq.~(\ref{F_SL}). The second correlation function is simplified as:
\bea
\la |\vp(i\om,q)|^2 \ra_B^{(2)} &&= {2 n_\rho^2 q^2 \ov (m'u_\rho^2)^2}~[\mathcal{D}(i\om,q)]^2~{1 \ov \beta}~\sum_{\nu}~\int {{dp \ov 2 \pi}}~\nu^2 (\nu-\om)^2~\mathcal{D}(i\nu,p)\mathcal{D}(i\nu-i\om,p-q)
\nonum \\
&&+ {8 n_\rho^2 q \om \ov (m'u_\rho^2)^2}~[\mathcal{D}(i\om,q)]^2~{1 \ov \beta}~\sum_{\nu}~\int {{dp \ov 2 \pi}}~p \nu (\nu-\om)^2~\mathcal{D}(i\nu,p)\mathcal{D}(i\nu-i\om,p-q)
\nonum \\
&&+ {4 n_\rho^2 \om^2 \ov (m'u_\rho^2)^2}~[\mathcal{D}(i\om,q)]^2~{1 \ov \beta}~\sum_{\nu}~\int {{dp \ov 2 \pi}}~p^2 (\nu-\om)^2~\mathcal{D}(i\nu,p)\mathcal{D}(i\nu-i\om,p-q)
\nonum \\
&&+ {4 n_\rho^2 \om^2 \ov (m'u_\rho^2)^2}~[\mathcal{D}(i\om,q)]^2~{1 \ov \beta}~\sum_{\nu}~\int {{dp \ov 2 \pi}}~p(p-q)\nu (\nu-\om)~\mathcal{D}(i\nu,p)\mathcal{D}(i\nu-i\om,p-q).
\nonum
\eea
All terms contribute to the on-shell part of the correlation function but only the first contributes to the off-shell part. The result for the self-energy part reads:
\bea
\Im \Sigma_B^R(\om,q) &&= {3 n_\rho^2 \ov 8 u_\rho}~\left( {v \ov u_\rho} \right)^2~\left( q^2 \ov m' \right)^2~u_\rho q~\left[ \delta[\om - u_\rho q] - \delta[\om + u_\rho q] \right]
\nonum \\
&&+ {n_\rho^2 \ov 16 u_\rho}~\left( {v \ov u_\rho} \right)^2~\left( q \ov m' u_\rho \right)^2~\left[ \om^2 - (u_\rho q)^2 \right]~\mathcal{F}_\rho(\om,q).
\eea
Finally, the third correlation function is simplified as:
\bea
\la |\vp(i\om,q)|^2 \ra_C^{(2)} &&= -{12 n_\rho u_\rho^2 q^2 \ov (m'u_\rho^2)^2}~[\mathcal{D}(i\om,q)]^2~{1 \ov \beta}~\sum_{\nu}~\int {{dp \ov 2 \pi}}~p(p-q) \nu (\nu-\om)~\mathcal{D}(i\nu,p)\mathcal{D}(i\nu-i\om,p-q)
\nonum \\
&&- {24 n_\rho u_\rho^2 q \om \ov (m'u_\rho^2)^2}~[\mathcal{D}(i\om,q)]^2~{1 \ov \beta}~\sum_{\nu}~\int {{dp \ov 2 \pi}}~p^2 (p-q) (\nu-\om)~\mathcal{D}(i\nu,p)\mathcal{D}(i\nu-i\om,p-q).
\nonum
\eea
Again, only the first term (which carries the $q^2$ factor) contributes to the off-shell part whereas both contribute to the on-shell part. The corresponding self-energy part reads:
\bea
\Im \Sigma_C^R(\om,q) &&= {3 n_\rho \ov 4 u_\rho}~\left( {v \ov u_\rho} \right)^2~\left( q^2 \ov m' \right)^2~u_\rho q~\left[ \delta[\om - u_\rho q] - \delta[\om + u_\rho q] \right]
\nonum \\
&&- {3 n_\rho \ov 8 u_\rho}~\left( {v \ov u_\rho} \right)^2~\left( q \ov m' u_\rho \right)^2~\left[ \om^2 - (u_\rho q)^2 \right]~\mathcal{F}_\rho(\om,q).
\eea
Adding the above self-energy parts the factors $(1+n_\rho)^2 = (1+3/\gamma_\rho^2)^2$ and $(3-n_\rho)^2 = 9(1 - 1/\gamma_\rho^2)^2$ appear and the self-energy is given by Eq.~(\ref{SE_SF}).

\subsection{Second-order perturbation in curvature: $\vp,\theta-$representation}
\label{A:SL2}

Expanding the perturbation up to second order yields three contributions:
\bea
&&\la |\vp(i\om,q)|^2 \ra^{(2)} = {1 \ov 2 m'^2}~{1 \ov \beta^6}~\sum_{\om_1,\om_2,\om_3,\om_1',\om_2',\om_3'}~\int {{dq_1 dq_2 dq_3 \ov (2 \pi)^3} {dq_1' dq_2' dq_3' \ov (2 \pi)^3}}~\delta[1+2+3] \delta[1'+2'+3']
\nonum \\
&& q_1 q_2 q_3 q_1' q_2' q_3'~\la  \vp_{\om}(q) \vp_{-\om}(-q) \vp_{\om_1}(q_1) \vp_{\om_2}(q_2) \vp_{\om_3}(q_3) \vp_{\om_1'}(q_1') \vp_{\om_2'}(q_2') \vp_{\om_3'}(q_3') \ra_0,
\nonum \\
&&+ {9 \ov 2 m'^2}~{1 \ov \beta^6}~\sum_{\om_1,\om_2,\om_3,\om_1',\om_2',\om_3'}~\int {{dq_1 dq_2 dq_3 \ov (2 \pi)^3} {dq_1' dq_2' dq_3' \ov (2 \pi)^3}}~\delta[1+2+3] \delta[1'+2'+3']
\nonum \\
&& q_1 q_2 q_3 q_1' q_2' q_3'~\la  \vp_{\om}(q) \vp_{-\om}(-q) \vp_{\om_1}(q_1) \theta_{\om_2}(q_2) \theta_{\om_3}(q_3) \vp_{\om_1'}(q_1') \theta_{\om_2'}(q_2') \theta_{\om_3'}(q_3') \ra_0,
\nonum \\
&&+ {3 \ov m'^2}~{1 \ov \beta^6}~\sum_{\om_1,\om_2,\om_3,\om_1',\om_2',\om_3'}~\int {{dq_1 dq_2 dq_3 \ov (2 \pi)^3} {dq_1' dq_2' dq_3' \ov (2 \pi)^3}}~\delta[1+2+3] \delta[1'+2'+3']
\nonum \\
&& q_1 q_2 q_3 q_1' q_2' q_3' ~\la  \vp_{\om}(q) \vp_{-\om}(-q) \vp_{\om_1}(q_1) \theta_{\om_2}(q_2) \theta_{\om_3}(q_3) \vp_{\om_1'}(q_1') \vp_{\om_2'}(q_2') \vp_{\om_3'}(q_3') \ra_0.
\nonum
\eea
The first contribution is exactly the same as the $A-$term of the previous paragraph.

Wick ordering the next contributions we classify the diagrams according to the number of $\theta-$wings, see Fig.~\ref{Diagrams}. Diagrams with no  $\theta-$wing are labeled as SLA (SpinLess A). SLA1 corresponds to the previous A term. The next terms read:
\bea
&&\mathcal{D}^{(2)}_{A2 \vp \vp}(i\om,q) = -18~[\mathcal{D}^{(0)}_{\vp \vp}(i\om,q)]^2~\left({q \ov m'} \right)^2~{1 \ov \beta}~\sum_\nu~\int {{dp \ov 2 \pi}}~p^2(p-q)^2~\mathcal{D}^{(0)}_{\theta \theta}(i\nu,p)~\mathcal{D}^{(0)}_{\theta \theta}(i\nu-i\om,p-q)
\nonum \\
&&\mathcal{D}^{(2)}_{A3 \vp \vp}(i\om,q) = - 36~[\mathcal{D}^{(0)}_{\vp \vp}(i\om,q)]^2~\left({q \ov m'} \right)^2~{1 \ov \beta}~\sum_\nu~\int {{dp \ov 2 \pi}}~p^2(p-q)^2~\mathcal{D}^{(0)}_{\vp \theta}(i\nu,p)~\mathcal{D}^{(0)}_{\theta \vp}(i\nu - i\om,p-q).
\nonum
\eea
Their self-energy parts reads:
\bea
\Im \Sigma_{A1}^R(\om,q) &&= {3 \gamma_\rho^2 \ov 8 u_\rho}~\left( q^2 \ov m' \right)^2~u_\rho q~\left[ \delta[\om - u_\rho q] - \delta[\om + u_\rho q] \right]
\nonum \\
&&+ {9 \gamma_\rho^2 \ov 16 u_\rho}~\left( q \ov m' u_\rho \right)^2~\left[ \om^2 - (u_\rho q)^2 \right]~{\mathcal{F}}_\rho(\om,q),
\nonum \\
\Im \Sigma_{A2}^R(\om,q) &&= { 3 \ov 8 u_\rho \gamma_\rho^2}~\left( q^2 \ov m' \right)^2~u_\rho q~\left[ \delta[\om - u_\rho q] - \delta[\om + u_\rho q] \right]
\nonum \\
&&+ {9 \ov 16 u_\rho \gamma_\rho^2}~\left( q \ov m' u_\rho \right)^2~\left[ \om^2 - (u_\rho q)^2 \right]~\mathcal{F}_\rho(\om,q),
\nonum \\
\Im \Sigma_{A3}^R(\om,q) &&= { 3 \ov 4 u_\rho}~\left( q^2 \ov m' \right)^2~u_\rho q~\left[ \delta[\om - u_\rho q] - \delta[\om + u_\rho q] \right]
\nonum \\
&&- {9 \ov 8 u_\rho}~\left( q \ov m' u_\rho \right)^2~\left[ \om^2 - (u_\rho q)^2 \right]~\mathcal{F}_\rho(\om,q).
\nonum
\eea
Diagrams with one $\theta-$wing (SLB) read:
\bea
&&\mathcal{D}^{(2)}_{B1 \vp \vp}(i\om,q) = 72~\mathcal{D}^{(0)}_{\vp \theta}(i\om,q)~\mathcal{D}^{(0)}_{\vp \vp}(i\om,q)~\left({q \ov m'} \right)^2~{1 \ov \beta}~\sum_\nu~\int {{dp \ov 2 \pi}}~p^2(p-q)^2~\mathcal{D}^{(0)}_{\vp \theta}(i\nu,p)~\mathcal{D}^{(0)}_{\theta \theta}(i\nu - i\om,p-q)
\nonum \\
&&\mathcal{D}^{(2)}_{B2 \vp \vp}(i\om,q) = - 72~\mathcal{D}^{(0)}_{\vp \theta}(i\om,q)~\mathcal{D}^{(0)}_{\vp \vp}(i\om,q)~\left({q \ov m'} \right)^2~{1 \ov \beta}~\sum_\nu~\int {{dp \ov 2 \pi}}~p^2(p-q)^2~\mathcal{D}^{(0)}_{\vp \theta}(i\nu,p)~\mathcal{D}^{(0)}_{\vp \vp}(i\nu - i\om,p-q).
\nonum
\eea
The corresponding self-energy parts read:
\bea
\Im \Sigma_{B1}^R(\om,q) &&= { 3 \ov 2 u_\rho \gamma_\rho^2}~\left( q^2 \ov m' \right)^2~u_\rho q~\left[ \delta[\om - u_\rho q] - \delta[\om + u_\rho q] \right],
\nonum \\
\Im \Sigma_{B2}^R(\om,q) &&= { 3 \ov 2 u_\rho}~\left( q^2 \ov m' \right)^2~u_\rho q~\left[ \delta[\om - u_\rho q] - \delta[\om + u_\rho q] \right].
\nonum
\eea
Diagrams with two $\theta-$wings (SLC) read:
\bea
&&\mathcal{D}^{(2)}_{C1 \vp \vp}(i\om,q) = 36~[\mathcal{D}^{(0)}_{\vp \theta}(i\om,q)]^2~\left({q \ov m'} \right)^2~{1 \ov \beta}~\sum_\nu~\int {{dp \ov 2 \pi}}~p^2(p-q)^2~\mathcal{D}^{(0)}_{\vp \vp}(i\nu,p)~\mathcal{D}^{(0)}_{\theta \theta}(i\nu - i\om,p-q)
\nonum \\
&&+36~[\mathcal{D}^{(0)}_{\vp \theta}(i\om,q)]^2~\left({q \ov m'} \right)^2~{1 \ov \beta}~\sum_\nu~\int {{dp \ov 2 \pi}}~p^2(p-q)^2~\mathcal{D}^{(0)}_{\vp \theta}(i\nu,p)~\mathcal{D}^{(0)}_{\vp \theta}(i\nu - i\om,p-q).
\nonum
\eea
The corresponding self-energy parts read:
\bea
\Im \Sigma_{C1}^R(\om,q) &&= { 3 \ov 4 u_\rho \gamma_\rho^2}~\left( q^2 \ov m' \right)^2~u_\rho q~\left[ \delta[\om - u_\rho q] - \delta[\om + u_\rho q] \right]
\nonum \\
&&+ {9 \ov 8 u_\rho \gamma_\rho^2}~\left( q \ov m' u_\rho \right)^2~\left[ \om^2 - (u_\rho q)^2 \right]~\mathcal{F}_\rho (\om,q),
\nonum \\
\Im \Sigma_{C2}^R(\om,q) &&= { 3 \ov 4 u_\rho \gamma_\rho^2}~\left( q^2 \ov m' \right)^2~u_\rho q~\left[ \delta[\om - u_\rho q] - \delta[\om + u_\rho q] \right]
\nonum \\
&&- {9 \ov 8 u_\rho \gamma_\rho^2}~\left( q \ov m' u_\rho \right)^2~\left[ \om^2 - (u_\rho q)^2 \right]~\mathcal{F}_\rho (\om,q).
\nonum
\eea
The total contribution of each set read:
\bea
&&\Im \Sigma_{A}^R(\om,q) = {3 \gamma_\rho^2 \ov 8 u_\rho}~\left[ 1 + {1 \ov \gamma_\rho^2} \right]^2~\left( q^2 \ov m' \right)^2~u_\rho q~\left[ \delta[\om - u_\rho q] - \delta[\om + u_\rho q] \right]
\nonum \\
&&+ {9 \gamma_\rho^2 \ov 16 u_\rho}~\left[ 1 - {1 \ov \gamma_\rho^2} \right]^2~\left( q \ov m' u_\rho \right)^2~\left[ \om^2 - (u_\rho q)^2 \right]~\mathcal{F}_\rho (\om,q)
\nonum \\
&&\Im \Sigma_{B}^R(\om,q) = {3 \ov 2 u_\rho}~\left[ 1 + {1 \ov \gamma_\rho^2} \right]~\left( q^2 \ov m' \right)^2~u_\rho q~\left[ \delta[\om - u_\rho q] - \delta[\om + u_\rho q] \right]
\nonum \\
&&\Im \Sigma_{C}^R(\om,q) = {3 \ov 2 u_\rho \gamma_\rho^2}~\left( q^2 \ov m' \right)^2~u_\rho q~\left[ \delta[\om - u_\rho q] - \delta[\om + u_\rho q] \right].
\eea
It is interesting to see that all diagrams contribute to the on-shell self-energy; however, the final expression for the tail comes only from the SLA set, {\it i.e.} those diagrams with no $\theta-$wings. The set SLB with a single $\theta-$wing does not have any off-shell contribution whereas the set SLC with two $\theta-$wings has off-shell contributions which cancel each-other. Notice that if the latter had a net contribution to the tails, this contribution would have higher powers of frequency which would violate the f-sum rule. In dimensionless units these additional contributions are: $\propto \om/ u_\rho q$ for SLB and $\propto (\om/ u_\rho q)^2$ for SLC due to the presence of a single $\theta-$wing and two $\theta-$wings. On-shell, {\it i.e.} for $\om = \pm u_\rho q$, such contributions are $\pm 1$ which yield the correct signs in front of the delta functions.

Adding all the above self-energies yields:
\bea
&&\Im \Sigma_\rho^R(\om,q) = {3 \gamma_\rho^2 \ov 8 u_\rho}~\left[ 1 + {3 \ov \gamma_\rho^2} \right]^2~\left( q^2 \ov m' \right)^2~u_\rho q~\left[ \delta[\om - u_\rho q] - \delta[\om + u_\rho q] \right]
\nonum \\
&&+ {9 \gamma_\rho^2 \ov 16 u_\rho}~\left[ 1 - {1 \ov \gamma_\rho^2} \right]^2~\left( q \ov m' u_\rho \right)^2~\left[ \om^2 - (u_\rho q)^2 \right]~\mathcal{F}_\rho(\om,q).
\eea
Substituting $m' = 6m/\sqrt{\pi}$ yields Eq.~(\ref{SE_SF}) which is precisely what has been found in the previous paragraph.


\section{Correlation functions in the spinful fermion case}
\label{A:SF}

\subsection{Charge structure factor}
\label{A:SF_CS}

With the help of the Lagrangian Eq.~(\ref{lagrangian_spinful}) the second-order contributions to the charge phase correlator have three contributions:
\bea
\la |\vp(i\om,q)|^2 \ra_A^{(2)} = {\mathrm{Similar~to~the~spinless~case}},
\nonum
\eea
which is identical to what is found in the spinless case,
\bea
&&\la |\vp(i\om,q)|^2 \ra_B^{(2)} =
\nonum \\
&&-18~[\mathcal{D}^{(0)}_{\vp \vp}(i\om,q)]^2~\left({q \ov m'} \right)^2~{1 \ov \beta}~\sum_\nu~\int {{dp \ov 2 \pi}}~p^2(p-q)^2~\mathcal{D}^{(0)}_{\sigma \sigma}(i\nu,p)~\mathcal{D}^{(0)}_{\sigma \sigma}(i\nu-i\om,p-q)
\nonum \\
&&-36~[\mathcal{D}^{(0)}_{\vp \vp}(i\om,q)]^2~\left({q \ov m'} \right)^2~{1 \ov \beta}~\sum_\nu~\int {{dp \ov 2 \pi}}~p^2(p-q)^2~\mathcal{D}^{(0)}_{\sigma \theta_\sigma}(i\nu,p)~\mathcal{D}^{(0)}_{\sigma \theta_\sigma}(i\nu-i\om,p-q)
\nonum \\
&&-18~[\mathcal{D}^{(0)}_{\vp \vp}(i\om,q)]^2~\left({q \ov m'} \right)^2~{1 \ov \beta}~\sum_\nu~\int {{dp \ov 2 \pi}}~p^2(p-q)^2~\mathcal{D}^{(0)}_{\theta_\sigma \theta_\sigma}(i\nu,p)~\mathcal{D}^{(0)}_{\theta_\sigma \theta_\sigma}(i\nu-i\om,p-q),
\nonum
\eea
\bea
&&\la |\vp(i\om,q)|^2 \ra_C^{(2)} =
\nonum \\
&&-72~\mathcal{D}^{(0)}_{\vp \vp}(i\om,q)~\mathcal{D}^{(0)}_{\vp \theta_\rho}(i\om,q)~\left({q \ov m'} \right)^2~{1 \ov \beta}~\sum_\nu~\int {{dp \ov 2 \pi}}~p^2(p-q)^2~\mathcal{D}^{(0)}_{\theta_\sigma \sigma}(i\nu,p)~\mathcal{D}^{(0)}_{\theta_\sigma \theta_\sigma}(i\nu-i\om,p-q)
\nonum \\
&&-72~\mathcal{D}^{(0)}_{\vp \vp}(i\om,q)~\mathcal{D}^{(0)}_{\vp \theta_\rho}(i\om,q)~\left({q \ov m'} \right)^2~{1 \ov \beta}~\sum_\nu~\int {{dp \ov 2 \pi}}~p^2(p-q)^2~\mathcal{D}^{(0)}_{\sigma \sigma}(i\nu,p)~\mathcal{D}^{(0)}_{\sigma \theta_\sigma}(i\nu-i\om,p-q)
\nonum
\eea
\bea
&&\la |\vp(i\om,q)|^2 \ra_D^{(2)} =
\nonum \\
&&-36~[\mathcal{D}^{(0)}_{\vp \theta_\rho}(i\om,q)]^2~\left({q \ov m'} \right)^2~{1 \ov \beta}~\sum_\nu~\int {{dp \ov 2 \pi}}~p^2(p-q)^2~\mathcal{D}^{(0)}_{\sigma \sigma}(i\nu,p)~\mathcal{D}^{(0)}_{\theta_\sigma \theta_\sigma}(i\nu-i\om,p-q)
\nonum \\
&&-36~[\mathcal{D}^{(0)}_{\vp \theta_\rho}(i\om,q)]^2~\left({q \ov m'} \right)^2~{1 \ov \beta}~\sum_\nu~\int {{dp \ov 2 \pi}}~p^2(p-q)^2~\mathcal{D}^{(0)}_{\sigma \theta_\sigma}(i\nu,p)~\mathcal{D}^{(0)}_{\theta_\sigma \sigma}(i\nu-i\om,p-q).
\nonum
\eea
Based on the similarity of the calculation with the spinless case we directly give the expression of the various self-energy parts:
\bea
&&\Im \Sigma_B^R(\om,q) = {3 \gamma_\sigma^2 \ov 8 u_\sigma}~\left( 1 + {1 \ov \gamma_\sigma^2} \right)^2~\left( q^2 \ov m' \right)^2~u_\sigma q~\left[ \delta[\om - u_\sigma q] - \delta[\om + u_\sigma q] \right]
\nonum \\
&&+ {9 \gamma_\sigma^2 \ov 16 u_\sigma}~\left( 1 - {1 \ov \gamma_\sigma^2} \right)^2~\left( q \ov m' u_\sigma \right)^2~\left[ \om^2 - (u_\sigma q)^2 \right]~\mathcal{F}_\sigma(\om,q).
\nonum \\
&&\Im \Sigma_{C}^R(\om,q) = {3 \gamma_\sigma \ov 2 u_\sigma \gamma_\rho}~\left[ 1 + {1 \ov \gamma_\sigma^2} \right]~\left( q^2 \ov m' \right)^2~u_\sigma q~\left[ \delta[\om - u_\sigma q] - \delta[\om + u_\sigma q] \right]
\nonum \\
&&\Im \Sigma_{D}^R(\om,q) = {3 \ov 2 u_\sigma \gamma_\rho^2}~\left( q^2 \ov m' \right)^2~u_\sigma q~\left[ \delta[\om - u_\sigma q] - \delta[\om + u_\sigma q] \right],
\eea
where $\mathcal{F}_\sigma$ is given by Eq.~(\ref{F_sigma}). Notice that these expressions satisfy the basic properties of the diagrams with zero, one or two $\theta-$wings.

Adding all the above self-energies and substituting $m' = 6m/\sqrt{\pi}$ yields:
\bea
&&\Im \Sigma_\rho^R(\om,q) = {\gamma_\sigma^2 \pi \ov 96 u_\sigma}~\left[ 1 + {1 \ov \gamma_\sigma^2} + {2 \ov \gamma_\sigma \gamma_\rho} \right]^2~\left( q^2 \ov m \right)^2~u_\sigma q~\left[ \delta[\om - u_\sigma q] - \delta[\om + u_\sigma q] \right]
\nonum \\
&&+ {\gamma_\sigma^2 \pi \ov 64 u_\sigma}~\left[ 1 - {1 \ov \gamma_\sigma^2} \right]^2~\left( q \ov m u_\sigma \right)^2~\left[ \om^2 - (u_\sigma q)^2 \right]~\mathcal{F}_\sigma(\om,q).
\eea
Adding the part arising from the charge degrees of freedom, the total self-energy of interacting spinful fermions is given by Eq.~(\ref{spinful_SC}).


\subsection{Spin structure factor}
\label{A:SF_SS}

With the help of the Lagrangian Eq.~(\ref{lagrangian_spinful}) the second-order contributions to the self-energy part of the spin structure factor read:
\bea
&&\Sigma_A^{(2)}(i\om,q) =
\nonum \\
&&36~\left({q \ov m'} \right)^2~{1 \ov \beta}~\sum_\nu~\int {{dp \ov 2 \pi}}~p^2(p-q)^2~\mathcal{D}^{(0)}_{\vp \vp}(i\nu,p)~\mathcal{D}^{(0)}_{\sigma \sigma}(i\nu-i\om,p-q)
\nonum \\
&&36~\left({q \ov m'} \right)^2~{1 \ov \beta}~\sum_\nu~\int {{dp \ov 2 \pi}}~p^2(p-q)^2~\mathcal{D}^{(0)}_{\theta_\vp \theta_\vp}(i\nu,p)~\mathcal{D}^{(0)}_{\theta_\sigma \theta_\sigma}(i\nu-i\om,p-q)
\nonum \\
&&72~\left({q \ov m'} \right)^2~{1 \ov \beta}~\sum_\nu~\int {{dp \ov 2 \pi}}~p^2(p-q)^2~\mathcal{D}^{(0)}_{\vp \theta_\vp}(i\nu,p)~\mathcal{D}^{(0)}_{\sigma \theta_\sigma}(i\nu-i\om,p-q),
\nonum
\eea
\bea
&&\Sigma_B^{(2)}(i\om,q) =
\nonum \\
&&{72 \ov \gamma_\sigma}~{i \om \ov u_\sigma q}~\left({q \ov m'} \right)^2~{1 \ov \beta}~\sum_\nu~\int {{dp \ov 2 \pi}}~p^2(p-q)^2~\mathcal{D}^{(0)}_{\vp \vp}(i\nu,p)~\mathcal{D}^{(0)}_{\sigma \theta_\sigma}(i\nu-i\om,p-q)
\nonum \\
&&{72 \ov \gamma_\sigma}~{i \om \ov u_\sigma q}~\left({q \ov m'} \right)^2~{1 \ov \beta}~\sum_\nu~\int {{dp \ov 2 \pi}}~p^2(p-q)^2~\mathcal{D}^{(0)}_{\theta_\vp \theta_\vp}(i\nu,p)~\mathcal{D}^{(0)}_{\sigma \theta_\sigma}(i\nu-i\om,p-q)
\nonum \\
&&{72 \ov \gamma_\sigma}~{i \om \ov u_\sigma q}~\left({q \ov m'} \right)^2~{1 \ov \beta}~\sum_\nu~\int {{dp \ov 2 \pi}}~p^2(p-q)^2~\mathcal{D}^{(0)}_{\vp \theta_\vp}(i\nu,p)~\mathcal{D}^{(0)}_{\sigma \sigma}(i\nu-i\om,p-q),
\nonum \\
&&{72 \ov \gamma_\sigma}~{i \om \ov u_\sigma q}~\left({q \ov m'} \right)^2~{1 \ov \beta}~\sum_\nu~\int {{dp \ov 2 \pi}}~p^2(p-q)^2~\mathcal{D}^{(0)}_{\vp \theta_\vp}(i\nu,p)~\mathcal{D}^{(0)}_{\theta_\sigma \theta_\sigma}(i\nu-i\om,p-q),
\nonum
\eea
\bea
&&\Sigma_C^{(2)}(i\om,q) =
\nonum \\
&&{36 \ov \gamma_\sigma^2}~\left({i \om \ov u_\sigma q} \right)^2~\left({q \ov m'} \right)^2~{1 \ov \beta}~\sum_\nu~\int {{dp \ov 2 \pi}}~p^2(p-q)^2~\mathcal{D}^{(0)}_{\vp \vp}(i\nu,p)~\mathcal{D}^{(0)}_{\theta_\sigma \theta_\sigma}(i\nu-i\om,p-q)
\nonum \\
&&{36 \ov \gamma_\sigma^2}~\left({i \om \ov u_\sigma q} \right)^2~\left({q \ov m'} \right)^2~{1 \ov \beta}~\sum_\nu~\int {{dp \ov 2 \pi}}~p^2(p-q)^2~\mathcal{D}^{(0)}_{\theta_\vp \theta_\vp}(i\nu,p)~\mathcal{D}^{(0)}_{\sigma \sigma}(i\nu-i\om,p-q)
\nonum \\
&&{72 \ov \gamma_\sigma^2}~\left({i \om \ov u_\sigma q} \right)^2~\left({q \ov m'} \right)^2~{1 \ov \beta}~\sum_\nu~\int {{dp \ov 2 \pi}}~p^2(p-q)^2~\mathcal{D}^{(0)}_{\vp \theta_\vp}(i\nu,p)~\mathcal{D}^{(0)}_{\sigma \theta_\sigma}(i\nu-i\om,p-q).
\nonum
\eea
The corresponding imaginary parts read:
\bea
&& \Im \Sigma_A^R(\om,q) = {9 \ov 2 u_\sigma}~\gamma_\rho \gamma_\sigma~\left[ 1 + {1 \ov \gamma_\rho \gamma_\sigma} \right]^2~\left({q^2 \ov m'} \right)^2~u_\sigma q~{1 \ov |u_\rho - u_\sigma|^3 q^3}
\nonum \\
&& \left \{ [\om - u_\sigma q]~[\om - u_\rho q]~\left[ n_B \left[ {u_\rho [\om - u_\sigma q] \ov u_\rho - u_\sigma}\right] - n_B \left[ {u_\sigma [\om - u_\rho q] \ov u_\rho - u_\sigma}\right] \right] \right.
\nonum \\
&& \left. + [\om + u_\sigma q]~[\om + u_\rho q]~\left[ n_B \left[ {u_\rho [\om + u_\sigma q] \ov u_\rho - u_\sigma}\right] - n_B \left[ {u_\sigma [\om + u_\rho q] \ov u_\rho - u_\sigma}\right] \right] \right \}
\nonum \\
&& + {9 \ov 2}~\gamma_\rho \gamma_\sigma~\left[ 1 - {1 \ov \gamma_\rho \gamma_\sigma} \right]^2~\left({q \ov m'} \right)^2~{1 \ov |u_\rho + u_\sigma|^3 q^3}
\nonum \\
&& \left \{ [\om + u_\sigma q]~[\om - u_\rho q]~\left[ n_B \left[ - {u_\sigma [\om - u_\rho q] \ov u_\rho + u_\sigma}\right] - n_B \left[ {u_\rho [\om + u_\sigma q] \ov u_\rho + u_\sigma}\right] \right] \right.
\nonum \\
&& \left. + [\om - u_\sigma q]~[\om + u_\rho q]~\left[ n_B \left[ - {u_\sigma [\om + u_\rho q] \ov u_\rho + u_\sigma} \right] - n_B \left[ {u_\rho [\om - u_\sigma q] \ov u_\rho + u_\sigma} \right] \right] \right \},
\eea
\bea
&& \Im \Sigma_B^R(\om,q) = {9 \ov u_\sigma \gamma_\sigma}~\left[ \gamma_\rho + \gamma_\sigma + {1 \ov \gamma_\rho} + {1 \ov \gamma_\sigma} \right]~{ \om \ov u_\sigma q}~\left({q^2 \ov m'} \right)^2~u_\sigma q~{1 \ov |u_\rho - u_\sigma|^3 q^3}
\nonum \\
&& \left \{ [\om - u_\sigma q]~[\om - u_\rho q]~\left[ n_B \left[ {u_\rho [\om - u_\sigma q] \ov u_\rho - u_\sigma}\right] - n_B \left[ {u_\sigma [\om - u_\rho q] \ov u_\rho - u_\sigma}\right] \right] \right.
\nonum \\
&& \left. - [\om + u_\sigma q]~[\om + u_\rho q]~\left[ n_B \left[ {u_\rho [\om + u_\sigma q] \ov u_\rho - u_\sigma}\right] - n_B \left[ {u_\sigma [\om + u_\rho q] \ov u_\rho - u_\sigma}\right] \right] \right \}
\nonum \\
&& - {9 \ov \gamma_\sigma}~\left[ \gamma_\rho - \gamma_\sigma + {1 \ov \gamma_\rho} - {1 \ov \gamma_\sigma} \right]~{ \om \ov u_\sigma q}~\left({q \ov m'} \right)^2~{1 \ov |u_\rho + u_\sigma|^3}
\nonum \\
&& \left \{ [\om + u_\sigma q]~[\om - u_\rho q]~\left[ n_B \left[ - {u_\sigma [\om - u_\rho q] \ov u_\rho + u_\sigma}\right] - n_B \left[ {u_\rho [\om + u_\sigma q] \ov u_\rho + u_\sigma}\right] \right] \right.
\nonum \\
&& \left. - [\om - u_\sigma q]~[\om + u_\rho q]~\left[ n_B \left[ - {u_\sigma [\om + u_\rho q] \ov u_\rho + u_\sigma} \right] - n_B \left[ {u_\rho [\om - u_\sigma q] \ov u_\rho + u_\sigma} \right] \right] \right \},
\eea
\bea
&& \Im \Sigma_C^R(\om,q) = {9 \ov 2 u_\sigma \gamma_\sigma \gamma_\rho}~\left[ 1 + {\gamma_\rho \ov \gamma_\sigma} \right]^2~\left[ { \om \ov u_\sigma q} \right]^2~\left({q^2 \ov m'} \right)^2~u_\sigma q~{1 \ov |u_\rho - u_\sigma|^3 q^3}
\nonum \\
&& \left \{ [\om - u_\sigma q]~[\om - u_\rho q]~\left[ n_B \left[ {u_\rho [\om - u_\sigma q] \ov u_\rho - u_\sigma}\right] - n_B \left[ {u_\sigma [\om - u_\rho q] \ov u_\rho - u_\sigma}\right] \right] \right.
\nonum \\
&& \left. + [\om + u_\sigma q]~[\om + u_\rho q]~\left[ n_B \left[ {u_\rho [\om + u_\sigma q] \ov u_\rho - u_\sigma}\right] - n_B \left[ {u_\sigma [\om + u_\rho q] \ov u_\rho - u_\sigma}\right] \right] \right \}
\nonum \\
&& + {9 \ov 2 \gamma_\sigma \gamma_\rho}~\left[ 1 - {\gamma_\rho \ov \gamma_\sigma} \right]^2~\left[{ \om \ov u_\sigma q}\right]^2~\left({q \ov m'} \right)^2~{1 \ov |u_\rho + u_\sigma|^3}
\nonum \\
&& \left \{ [\om + u_\sigma q]~[\om - u_\rho q]~\left[ n_B \left[ - {u_\sigma [\om - u_\rho q] \ov u_\rho + u_\sigma}\right] - n_B \left[ {u_\rho [\om + u_\sigma q] \ov u_\rho + u_\sigma}\right] \right] \right.
\nonum \\
&& \left. + [\om - u_\sigma q]~[\om + u_\rho q]~\left[ n_B \left[ - {u_\sigma [\om + u_\rho q] \ov u_\rho + u_\sigma} \right] - n_B \left[ {u_\rho [\om - u_\sigma q] \ov u_\rho + u_\sigma} \right] \right] \right \},
\eea
where $n_B$ is the Bose occupation function. Summing these contributions yields:
\bea
&& \Im \Sigma_\sigma^R(\om,q) = {9 \gamma_\rho \gamma_\sigma \ov 2 u_\sigma}~\left[ 1 + {1 \ov \gamma_\rho \gamma_\sigma} + {1 \ov \gamma_\rho \gamma_\sigma}~\left[ 1 + {\gamma_\rho \ov \gamma_\sigma}\right]~{\om \ov u_\sigma q} \right]^2~\left({q^2 \ov m'} \right)^2~u_\sigma q
\nonum \\
&& {[\om - u_\sigma q]~[\om - u_\rho q] \ov |u_\rho - u_\sigma|^3 q^3}~\left[ n_B \left[ {u_\rho [\om - u_\sigma q] \ov u_\rho - u_\sigma}\right] - n_B \left[ {u_\sigma [\om - u_\rho q] \ov u_\rho - u_\sigma}\right] \right]
\nonum \\
&& + {9 \gamma_\rho \gamma_\sigma \ov 2 u_\sigma}~\left[ 1 + {1 \ov \gamma_\rho \gamma_\sigma} - {1 \ov \gamma_\rho \gamma_\sigma}~\left[ 1 + {\gamma_\rho \ov \gamma_\sigma}\right]~{\om \ov u_\sigma q} \right]^2~\left({q^2 \ov m'} \right)^2~u_\sigma q
\nonum \\
&& {[\om + u_\sigma q]~[\om + u_\rho q] \ov |u_\rho - u_\sigma|^3 q^3}~\left[ n_B \left[ {u_\rho [\om + u_\sigma q] \ov u_\rho - u_\sigma}\right] - n_B \left[ {u_\sigma [\om + u_\rho q] \ov u_\rho - u_\sigma}\right] \right]
\nonum \\
&& + {9 \gamma_\rho \gamma_\sigma \ov 2 }~\left[ 1 - {1 \ov \gamma_\rho \gamma_\sigma} + {1 \ov \gamma_\rho \gamma_\sigma}~\left[ 1 - {\gamma_\rho \ov \gamma_\sigma} \right]~{\om \ov u_\sigma q} \right]^2~\left({q \ov m'} \right)^2~{1 \ov |u_\rho + u_\sigma|^3}
\nonum \\
&& [\om + u_\sigma q]~[\om - u_\rho q]~\left[ n_B \left[ - {u_\sigma [\om - u_\rho q] \ov u_\rho + u_\sigma}\right] - n_B \left[ {u_\rho [\om + u_\sigma q] \ov u_\rho + u_\sigma}\right] \right]
\nonum \\
&& + {9 \gamma_\rho \gamma_\sigma \ov 2}~\left[ 1 - {1 \ov \gamma_\rho \gamma_\sigma} - {1 \ov \gamma_\rho \gamma_\sigma}~\left[ 1 - {\gamma_\rho \ov \gamma_\sigma} \right]~{\om \ov u_\sigma q} \right]^2~\left({q \ov m'} \right)^2~{1 \ov |u_\rho + u_\sigma|^3}
\nonum \\
&& [\om - u_\sigma q]~[\om + u_\rho q]~\left[ n_B \left[ - {u_\sigma [\om + u_\rho q] \ov u_\rho + u_\sigma}\right] - n_B \left[ {u_\rho [\om - u_\sigma q] \ov u_\rho + u_\sigma}\right] \right].
\label{SE_SF_SS}
\eea
At $T=0$ and in the limit where spin and charge parameters are equal ($\gamma_\sigma = \gamma_\rho$, $u_\sigma = u_\rho$) we recover the result of Eq.~(\ref{SE_SF}) with an additional factor of $2$, {\it i.e.} the spin degeneracy. At $T=0$ and in the absence of backscattering ($\gamma_\sigma = 1$, $u_\sigma =v$), Eq.~(\ref{SE_SF_SS}) reduces to Eq.~(\ref{SE_SF_SS2}).

\end{widetext}



\begin{thebibliography}{99}
\bibitem{Note:fractional}
This agrees with our earlier result, derived in Ref.~[\onlinecite{ST}], that band-curvature yields sub-leading corrections to the electron self-energy: $\Im \Sigma \sim |\eps| + |\eps|^3/\eps_F^2$, where $\eps_F = mv^2$. The leading linear part is a signature that the electrons are not the stable quasi-particles of the system.
\bibitem{Note:NAB}
The model of spinful fermions is invariant under the large group: $U(1) \times SU(2)$, which has a non-Abelian component arising from the spin of the fermions. A rigorous procedure would require using the non-Abelian bosonization technique to treat the spin sector. This would yield the so-called $SU(2)$ level $1$ Sugawara Hamiltonian and the corresponding Wess-Zumino-Witten (WZW) Lagrangian, see the textbook Ref.~[\onlinecite{Book:GNT}], properly generalized to include the effect of curvature. A formulation of the problem within the WZW non-linear $\sigma-$model is thus possible. In the present study, we restrict ourselves to a simpler Abelian bosonization and, whenever necessary, mention the constraints arising from the $SU(2)$ symmetry.
\bibitem{Note:chargeDSF_Hubbard}
In the literature on the Hubbard model both of these peaks are known, {\it e.g.} in Ref.~[\onlinecite{Book:Essler}] see Fig. 7.29 for the particle-hole peak and Fig. 7.31 for the singlet peak (the latter is degenerate with the spin-triplet).
\end{thebibliography}
\end{document}